\documentclass[11pt]{article} 
\usepackage{amsmath} 
\usepackage{amssymb} 
\usepackage{graphicx} 
\usepackage{geometry}

\newcommand{\bsigma}{\mbox{\boldmath $\sigma$}} 
\newcommand{\izq}{\left\langle} 
\newcommand{\der}{\right\rangle_{\Psi}} 
 
\newcommand{\smprod}[2]{{\bf #1}{\bf #2}^\dagger} 
\newcommand{\sprod}[2]{{\bf #1} \cdot {\bf #2}} 

\begin{document}
\title{\bf The Blume-Emery-Griffiths neural network: dynamics for
arbitrary temperature} 
\author{D. Boll\'e$^a$, 
        J. Busquets Blanco$^a$, 
	G.M.Shim$^b$ 
	and T. Verbeiren$^a$}
\date{}
\maketitle
\begin{center}
$^a$ Instituut voor Theoretische Fysica, Katholieke Universiteit Leuven,\\
Celestijnenlaan 200 D, B-3001 Leuven, Belgium\\
$^b$ Department of Physics, Chungnam National 
University,\\Yuseong, Daejeon 305-764, R.O.~Korea

\end{center}

\begin{abstract} 
The parallel dynamics of the fully connected Blume-Emery-Griffiths neural
network model is studied for arbitrary temperature. By employing a
probabilistic signal-to-noise approach, a recursive scheme is found
determining the time evolution of the distribution of the local fields and,
hence, the evolution of the order parameters. A comparison of this approach is
made with the generating functional method, allowing to calculate any physical
relevant quantity as a function of time. Explicit analytic formula are given
in both methods for the first few time steps of the dynamics.  Up to the third
time step the results are identical. Some arguments are presented why beyond
the third time step the results differ for certain values of the model
parameters. Furthermore, fixed-point equations are derived in the stationary
limit.	Numerical simulations confirm our theoretical findings.
\end{abstract} 

%\begin{keyword} 
%parallel dynamics; fully connected networks; probabilistic approach,
%generating functional approach; thermal fields. 
%\PACS 87.10.+e \sep 64.60.Cn \sep 75.10.Hk \sep 02.50.-r
%\end{keyword} 

\section{Introduction} \label{sec:intro} 

Recently, we have studied the Blume-Emery-Griffiths (BEG) neural network
dynamics at zero temperature using the probabilistic signal-to-noise analysis
(SNA) \cite{jo}. The interest of such a study arises from the fact that the
BEG-model optimises the mutual information content for three-state networks
\cite{DK00},\cite{BV02}. In this way the best retrieval properties including,
e.g., the largest retrieval overlap, loading capacity, basin of attraction and
convergence time, are guaranteed in comparison with other three-state networks
studied in the literature \cite{BV03}. As has been pointed out in \cite{BV03}
the system allows for new stable stationary states, i.e. quadrupolar states,
when the temperature becomes high enough.  Therefore, it is interesting to
study the dynamics of the model at arbitrary temperatures. This can be done in
the framework of the SNA by introducing auxiliary thermal fields \cite{PZ90}
to express the stochastic dynamics within the gain function formulation of the
deterministic dynamics. The calculations are rather involved because the
feedback in these systems \cite{BKS90} causes the appearance of discrete
noise, besides the Gaussian one, involving the neurons at all previous time
steps. This prevents a closed-form solution of the dynamics although it allows
for a recursive scheme to find the order parameters, starting from the time
evolution of the distribution of the local fields. Hence, it seems worthwhile
to also apply the generating functional approach (GFA) \cite{SMR73},
\cite{D78} to solve this feedback dynamics. This method enables us to find all
relevant physical order parameters at any time step via the derivatives of the
generating functional. 

Both methods are then compared for the first few time steps and the results
are analytically shown to be identical up to the third time step. Numerical
simulations are performed for systems up to $7000$ neurons and they confirm
the analytical results we have obtained.  Beyond the third time step it is
found numerically that the SNA approach certainly leads to different results
than those obtained through simulations for values of the model parameters
corresponding to spin-glass behaviour of the system. This is traced back to
the treatment of the residual overlap representing the influence of the
non-condensed patterns in the time evolution of the system. Finally, as in the
$T=0$ case, we derive fixed-point equations in the stationary limit and
compare them with existing results within a replica symmetric thermodynamic
approach \cite{BV03}. 

The paper is organized as follows. In Section~2 we introduce the BEG model and
the order parameters of interest. Section~3 is devoted to the solution of the
dynamics of the model with the SNA. The first few time steps are obtained for
the main overlap, the neural activity, the activity overlap and the variance
of the residual overlaps. Moreover, the stationary limit of the dynamics is
discussed. Furthermore, numerical simulations of this dynamics are presented
for the first three time steps showing agreement with the theoretical
findings.  Section~4 discusses the GFA approach to the BEG dynamics. In
Section 5 we compare both methods analytically up to the third time step and
numerically beyond that.  Some concluding remarks are given in Section~6. All
technical details are refered to Appendices.

\section{The BEG model} 
\label{sec:mod} 
Consider a neural network consisting of $N$ neurons which can
take values $\{\sigma_i| i=1,\ldots, N\}$ from the discrete set $
\mathcal{S}=\{s_k|\,k=1,2,3\}\equiv \lbrace -1,0,+1 \rbrace $. The
$p=\alpha N$ patterns to be stored in 
this network are supposed to be a collection of independent and
identically distributed random variables 
(i.i.d.r.v.), $\{\xi_i^\mu |\mu =1,\ldots,p ; i=1, \ldots, N\}$
with a probability distribution 
\begin{equation} 
p(\xi_{i}^{\mu})=\frac{a}{2}\delta(\xi_{i}^{\mu}-1)
+\frac{a}{2}\delta(\xi_{i}^{\mu}+1)+(1-a)\delta(\xi_{i}^{\mu})
\end{equation} 
with $a$ the activity of the patterns so that 
\begin{equation} 
  \lim_{N\rightarrow\infty}\frac{1}{N}\sum_{i}^N(\xi_{i}^{\mu})^2 =
  a. 
\end{equation} 
Given the network configuration at time $t$,
${\bsigma}_N(t)\equiv\{\sigma_j(t)| j=1,\ldots,N \}$, the neurons 
are updated according to the stochastic parallel spin-flip dynamics
defined by the transition probabilities 
\begin{equation} 
   \mbox{Prob} \left(\sigma_i(t+1) = s_k \in \mathcal{S}|
   \bsigma_N(t) \right) = 
	\frac 
	{\exp \{- \beta \epsilon_i[s_k|\bsigma_N(t)] \} } 
	{\sum_{s \in \mathcal{S}} \exp \{- \beta \epsilon_i 
				   [s|\bsigma_N(t)]\} }\,. 
\label{eq:trans} 
\end{equation} 
The configuration $\bsigma_N(0)$ is chosen as input. The energy
potential $\epsilon_i[s|{\bsigma}_N(t)]$ is 
defined by 
\begin{equation} 
	  \epsilon_i[s|{\bsigma}_N(t)] =
	  -sh_i({\bsigma}_N(t))-s^2\theta_i({\bsigma}_N(t)) 
					    \,, 
  \label{eq:energy} 
\end{equation} 
where the following local fields in neuron $i$ carry all the
information 
\begin{equation} 
	\label{eq:h} 
      h_i(\bsigma_N(t))=\sum_{j \neq i} J_{ij}\sigma_j(t), \quad 
      \theta_i(\bsigma_N(t))=\sum_{j\neq i}K_{ij}\sigma_{j}^{2}(t)
      \, . 
\end{equation} 
The synaptic couplings $J_{ij}$ and $K_{ij}$ are of the Hebb-type 
\begin{equation} 
J_{ij}=\frac{1}{a^{2}N}\sum_{\mu=1}^{p}\xi_{i}^{\mu}\xi_{j}^{\mu},\qquad
K_{ij}=\frac{1}{N}\sum_{\mu=1}^{p}\eta_{i}^{\mu}\eta_{j}^{\mu} 
\label{hebb} 
\end{equation} 
with 
\begin{equation} 
\eta_{i}^{\mu}= \frac{1}{\tilde{a}} ((\xi_{i}^{\mu})^{2}-a), \quad 
\tilde{a}= a(1-a). 
\end{equation} 
The order parameters relevant for the present discussion of this
model are 
\begin{equation} 
	\label{eq:mdef} 
	m_N^\mu(t)=\frac{1}{aN} 
		\sum_{i}\xi_i^\mu\sigma_i(t), 
		\quad 
       l_N^\mu(t)=\frac{1}{N}\sum_{i}\eta_i^\mu\sigma_{i}^{2}(t), 
		    \quad 
       q_N(t)=\frac{1}{N}\sum_{i}\sigma_{i}^{2}(t) 
\end{equation} 
where $m_N^\mu$ is the retrieval overlap, $l_N^\mu(t)$ related to
the activity overlap, and $q_N(t)$ the neural activity \cite{DK00},
\cite{BV02}.

\section{The SNA for arbitrary temperatures} 
\label{sec:prob} 

\subsection{Recursive dynamical scheme} 
In \cite{jo} we discussed the SNA for the BEG model at temperature
$T=0$. In order to 
generalize this approach to finite temperatures we introduce
auxiliary thermal fields \cite{PZ90} in order to 
express the stochastic dynamics within the gain function
formulation of the deterministic dynamics. Consider
for each $i$ and $t$ the auxiliary thermal fields,
$\mathbf{\Psi}_i(t)=\{\psi_{i}^{l}(t)|\,l=1,2,3\}$, which are
i.i.d.r.v. with respect to $i$ and $t$, such that \cite{BSV94} 
\begin{multline}
\mbox{Prob}
  [\sigma_{i}(t+1)=  s_{k} \in \mathcal{S}|{\bsigma}_N(t)]=  \\ 
  \left[\prod_{l\neq k} 
  \Theta(\epsilon_{i}[s_{l}|\bsigma_N(t)]-\psi_{i}^{l}(t+1)- 
    \epsilon_{i}[s_{k}|\bsigma_N(t)]+\psi_{i}^{k}(t+1))\right]_{\Psi}
\label{eq:transzero} 
\end{multline}
with $[\cdot]_{\Psi}$ the expectation value with respect
to ${\bf \Psi}_i(t)$. The joint probability density of the
$\mathbf{\Psi}_i(t)$ with respect to $l$ follows from comparing
(\ref{eq:trans})-(\ref{eq:energy}) with (\ref{eq:transzero}) and
the usual normalisation to $1$. The result is 
\begin{equation} 
{f}_{\beta}[\mathbf{\Psi}_i(t)]= 
\beta^{2}3!\frac{\exp{[-\beta\sum_{l=1}^{3}\psi_{i}^{l}(t)]}} 
{\{\sum_{l=1}^{3}\exp{[-\beta\psi_{i}^{l}(t)]}\}^3}
\delta\left(\sum_{l=1}^{3}\psi_{i}^{l}(t)\right)
\, . 
\label{density} 
\end{equation} 
To make the link between the stochastic dynamics and the
deterministic dynamics in a convenient way we observe 
that for each realization of the $\mathbf{\Psi}_i(t+1)$ with density
${f}_{\beta}(\cdot)$ the network state 
evolves according to the deterministic rule 
\begin{align} 
\sigma_{i}(t+1)& =\sum_{k=1}^{3}s_{k}\prod_{l\neq k} 
\Theta(\epsilon_{i}[s_{l}|\bsigma_N(t)]-\psi_{i}^{l}(t+1) 
     -\epsilon_{i}[s_{k}|\bsigma_N(t)]+\psi_{i}^{k}(t+1)) 
    \nonumber \\ 
&\equiv g_{\Psi}[h_{i}(\bsigma_{N}(t)),\theta_{i}(\bsigma_{N}(t))]
\, . \label{gainpsi} 
\end{align} 
In this way the problem of deriving recursion relations for $T
\neq 0$ becomes tractable. We start from the following initial
conditions 
\begin{eqnarray} 
&& E[\sigma_{i}(0)]=0, \quad Var[\sigma_{i}(0)]= q_{0}\,, 
	 \label{initial1} \\ 
&& E[\xi_{i}^{\mu}\sigma_{j}(0)]= 
   \delta_{i,j} \delta_{\mu,1}m_0^{1}a, \quad m_{0}^{1}>0, 
    \quad E[\eta_{i}^{\mu}\sigma_{j}^2(0)]= 
	      \delta_{i,j} \delta_{\mu,1}l_0^{1} \, . 
	  \label{initial2} 
\end{eqnarray} 
So we assume that the initial configuration is correlated with only
one (i.e. the condensed) stored pattern. 
By the law of large numbers (LLN) Eqs.(\ref{eq:mdef}) and
(\ref{initial1})-(\ref{initial2}) 
determine the order parameters $m_N^1(0),q_N^1(0)$ and $l_N^1(0)$
at $t=0$ in the limit $N \to \infty$. In the following we leave
out the pattern index $1$. 

For a general time step we find from Eq.(\ref{gainpsi}) and the
LLN in the limit $N \to \infty$ for the order 
parameters (eqs.(\ref{eq:mdef})) 
\begin{eqnarray} 
 \label{defm} 
m^{1}(t+1)&\stackrel{\mathcal Pr}{=}& 
  \frac{1}{a}\left[\left\langle\!\left\langle\xi_i 
	     g_{\mathbf{\Psi}}\big(h_{i}(t),\theta_{i}(t)\big) 
\right\rangle\!\right\rangle\right]_{\Psi} 
	     \\ \label{defq} 
q(t+1)&\stackrel{\mathcal Pr}{=}& 
   \left[ \left\langle\!\left\langle
   g_{\mathbf{\Psi}}^{2}\big(h_{i}(t),\theta_{i}(t) 
       \big)\right\rangle\!\right\rangle\right]_{\Psi} 
	   \\ \label{defl} 
l^{1}(t+1)&\stackrel{\mathcal
Pr}{=}&\left[\left\langle\!\left\langle\eta_i 
	 g_{\mathbf{\Psi}}^{2}\big(h_{i}(t),\theta_{i}(t)\big) 
\right\rangle\!\right\rangle\right]_{\Psi} 
\end{eqnarray} 
where $h_{i}(t)=\lim_{N\rightarrow\infty}h_{i}(\bsigma_N(t))$
(with an analogous formula for $\theta_{i}(t)$), and where 
the convergence is in probability. In the above the average,
denoted by $\left\langle \!\left\langle\cdot 
\right\rangle\!\right\rangle$, has to be taken over both the
distribution of the $\{\xi_i^{\mu}\}$ and the $\{\sigma_i(0)\}$. Note
that the $\{\sigma_i(0)\}$ is hidden in the local 
fields through the updating rule (\ref{gainpsi}). For
the Eqs. (\ref{defm})-(\ref{defl}) the average over the
thermal fields can be done explicitly by using the relations
(\ref{eq:transzero})-(\ref{density}) leading to 
\begin{eqnarray} 
 \label{defmt} 
m^{1}(t+1)&\stackrel{\mathcal Pr}{=}&
\frac{1}{a}\left\langle\!\left\langle 
    \xi_i V_{\beta}\big(h_{i}(t),\theta_{i}(t)\big) 
				\right\rangle\!\right\rangle 
	     \\ \label{defqt} 
q(t+1)&\stackrel{\mathcal Pr}{=}& 
	     \left\langle\!\left\langle 
		W_{\beta}\big(h_{i}(t),\theta_{i}(t) 
			     \big)\right\rangle\!\right\rangle 
	      \\ \label{deflt} 
l^{1}(t+1)&\stackrel{\mathcal Pr}{=}& 
      \left\langle\!\left\langle 
      \eta_i W_{\beta}\big(h_{i}(t),\theta_{i}(t)\big) 
			    \right\rangle\!\right\rangle 
\end{eqnarray} 
with 
\begin{align} 
V_{\beta}(h_{i}(t),\theta_{i}(t)) &= 
    \frac{\sinh(\beta h_{i}(t))} 
	      {\frac12\exp{\left(-\beta\theta_{i}(t)\right)} + 
	      \cosh(\beta h_{i}(t))} 
\\ 
W_{\beta}(h_{i}(t),\theta_{i}(t)) &= 
    \frac{\cosh(\beta h_{i}(t))} 
	    {\frac12\exp{\left(-\beta\theta_{i}(t)\right)}+\cosh(\beta
	    h_{i}(t))} \, . 
\end{align} 
In order to obtain the recursion relations for the local fields we
can then follow the derivation of the zero temperature 
case \cite{jo} (see Section 3) step by step. We do not repeat this 
calculation here but write down the final results 
\begin{align} 
h_{i}(t+1)&= 
   \frac{\xi_i}{a}m(t+1) 
	+\chi_{h}(t)\Big\{h_{i}(t)-\frac{1}{a}\xi_{i}m(t)+ 
			      \frac{\alpha}{a}\sigma_{i}(t)\Big\} 
	   +\mathcal{N}\Big(0,\frac{\alpha}{a^{2}}q(t+1)\Big) 
\label{hprob}\\ 
\theta_{i}(t+1)&= 
    \eta_{i}l(t+1) 
       +\chi_{\theta}(t)\Big\{\theta_{i}(t)-\eta_{i}l(t)+ 
		\frac{\alpha}{\tilde{a}}\sigma_{i}^{2}(t)\Big\} 
      +\mathcal{N}\Big(0,\frac{\alpha}{\tilde{a}^{2}}q(t+1)\Big)
      \, . 
\label{thetaprob} 
\end{align} 
From this it is clear that the local fields at time $t+1$ consist
out of a discrete part and a normally 
distributed part 
\begin{eqnarray} 
h_{i}(t)&=& 
      M_{i}(t)+\mathcal{N}\big(0, \alpha a D(t)\big) 
	\label{rec1} \\ 
\theta_{i}(t)&=& 
      L_{i}(t)+\mathcal{N}\big(0, \frac{\alpha}{\tilde{a}}E(t)\big) 
\end{eqnarray} 
where 
\begin{eqnarray} 
M_{i}(t+1)&=& 
   \chi_{h}(t)\Big[M_{i}(t)-\frac{\xi_{i}}{a}m(t)+ 
     \frac{\alpha}{a}\sigma_{i}(t)\Big]+\frac{\xi_{i}}{a}m(t+1)\\ 
L_{i}(t+1)&=& 
     \chi_{\theta}(t)\Big[L_{i}(t)-\eta_{i}l(t)+ 
    \frac{\alpha}{\tilde{a}}\sigma_{i}^{2}(t)\Big]+\eta_{i}l(t+1) 
    \label{rec4} 
\end{eqnarray} 
with $D(t)$ and $E(t)$ the variances of the residual overlaps
defined by 
\begin{equation} 
\label{rs} 
r^{\mu}(t)=\lim_{N\rightarrow\infty} 
      \frac{1}{a^2\sqrt{N}}\sum_{j}\xi_{j}^{\mu}\sigma_{j}(t), 
      \quad 
s^{\mu}(t)=\lim_{N\rightarrow\infty} 
      \frac{1}{\sqrt{N}}\sum_{j}\eta_{j}^{\mu}\sigma_{j}^{2}(t), 
      \quad \mu > 1\,, 
\end{equation} 
and taking into account the noise produced by the non-condensed
patterns in the local fields. They satisfy 
the recursion relations 
\begin{eqnarray} 
D(t+1)&=& 
   \frac{q(t+1)}{a^{3}} +\chi_{h}^{2}(t)D(t)+ 
       2\chi_{h}(t)\mbox{Cov}[\tilde{r}^{\mu}(t), r^{\mu}(t)] 
      \label{drec} \\ 
E(t+1)&=& 
     \frac{q(t+1)}{\tilde{a}}+\chi_{\theta}^{2}(t)E(t)+ 
     2\chi_{\theta}(t)\mbox{Cov}[\tilde{s}^{\mu}(t),s^{\mu}(t)]\,. 
     \label{erec} 
\end{eqnarray} 
In these expressions, the modified residual overlaps
$\tilde{r}^{\mu}, \tilde{s}^{\mu}, \mu>1$ are introduced because
in the residual overlaps the local fields strongly depend upon
$\xi_i^{\mu}$ and $\eta_i^{\mu}$ respectively. Therefore, the local
fields for finite $N$ have been modified as follows 
\begin{equation} 
 \tilde{h}_{N,i}^{\mu}(t)= h_{N,i}(t)- 
	   \frac{1}{\sqrt{N}} \xi_i^{\mu}r_N^{\mu}(t)\, , \quad 
 \tilde{\theta}_{N,i}^{\mu}(t)= \theta_{N,i}(t)- 
	   \frac{1}{\sqrt{N}} \eta_i^{\mu}s_N^{\mu}(t)\, , 
   \label{fieldmod} 
\end{equation} 
such that, employing the central limit theorem (CLT) 
\begin{eqnarray} 
 \tilde{r}^{\mu}(t)&=& 
     \lim_{N\rightarrow\infty}\frac{1}{a^2\sqrt{N}} 
     \sum_{j}\xi_{j}^{\mu}
     g_{\Psi}[\tilde{h}_{N,j}(t),\tilde{\theta}_{N,j}(t)] 
      \stackrel{\mathcal Pr}{=} 
	    \mathcal{N}\Big(0,\frac{q(t+1)}{a^{3}}\Big) 
		      \label{tilder}\\ 
\tilde{s}^{\mu}(t)&=& 
      \lim_{N\rightarrow\infty}\frac{1}{\sqrt{N}} 
  \sum_{j}\eta_{j}^{\mu}g^2_{\Psi}[\tilde{h}_{N,j}(t),\tilde{\theta}_{N,j}(t)]
      \stackrel{\mathcal Pr}{=} 
	    \mathcal{N}\Big(0,\frac{q(t+1)}{\tilde{a}}\Big) \,, 
\label{tildes} 
\end{eqnarray} 
where we have now assumed that the $\tilde{r}^{\mu}, \tilde{s}^{\mu},
\mu>1$ depend only weakly on $\xi_i^{\mu}$ and $\eta_i^{\mu}$
respectively. 
We remark that in the thermodynamic limit the density distributions
of the modified local fields $\tilde{h}_{i}(t)$ and 
$\tilde{\theta}_{i}(t)$ equal the density
distributions of the local fields $h_{i}(t)$ and $\theta_{i}(t)$ itself. 
The $\chi_{h}(t)$ and $\chi_{\theta}(t)$ are the susceptibilities
corresponding to the fields $h_i(t)$ and $\theta_i(t)$ 
\begin{eqnarray} 
\chi_{h}(t)&=&\frac{1}{a}\Big\langle\! \Big\langle 
     \int d\hat{h}\int d\hat{\theta} \,\, 
  \rho_{h_i(t)}(\hat{h}) \rho_{\theta_i(t)}(\hat{\theta}) 
     \left.\Big(\frac{ 
     {\partial V_{\beta}(h(t),\theta(t))}} 
     {{\partial h(t)}}\Big)\right|_{(\hat{h},\hat{\theta})} 
      \Big\rangle\! \Big\rangle\\ 
\chi_{\theta}(t)&=&\frac{1}{\tilde{a}}\Big\langle\! \Big\langle 
 \int d\hat{h}\int d\hat{\theta}\,\, 
  \rho_{h_i(t)}(\hat{h})\rho_{\theta_i(t)}(\hat{\theta}) 
      \left. \Big(\frac{ 
     {\partial W_{\beta}(h(t),\theta(t))}} 
   {{\partial \theta(t)}}\Big)\right|_{(\hat{h},\hat{\theta})} 
      \Big\rangle\! \Big\rangle \,. 
\end{eqnarray} 
In these expressions, $\rho_{h_i(t)}(x)$ and $\rho_{\theta_i(t)}(x)$
are the probability densities of the local fields 
\begin{equation} 
\begin{split} 
\rho_{h_{i}(t)}(z) 
&=\int\big(\prod_{s=0}^{t-2}dx_{s}dy_{s}\big)dx_{t}dy_{t} 
   \frac{1}{\sqrt{\det{(2\pi {\bf C}_{h})}}}\frac{1}{\sqrt{\det{(2\pi
   {\bf C}_{\theta})}}} \\ 
&\times\exp\big(-\frac{1}{2}{\bf x}{\bf C}_{h}^{-1}{\bf
x}^{\dagger}-\frac{1}{2}{\bf y}{\bf C}_{\theta}^{-1}{\bf
y}^{\dagger}\big)\,\, 
       \delta\big(z-M_{i}(t)-\sqrt{\alpha a D(t)}x_{t}\big) 
\end{split} 
\end{equation} 
\begin{equation} 
\begin{split} 
\rho_{\theta_{i}(t)}(z) 
&=\int\big(\prod_{s=0}^{t-2}dx_{s}dy_{s}\big)dx_{t}dy_{t} 
   \frac{1}{\sqrt{\det{(2\pi {\bf C}_{h})}}}\frac{1}{\sqrt{\det{(2\pi
   {\bf C}_{\theta})}}} \\ 
&\times\exp\big(-\frac{1}{2}{\bf x}{\bf C}_{h}^{-1}{\bf
x}^{\dagger}-\frac{1}{2}{\bf y}{\bf C}_{\theta}^{-1}{\bf
y}^{\dagger}\big)\,\, 
   \delta\big(z-L_{i}(t)-\sqrt{\frac{\alpha}{\tilde a}E(t)}y_{t}\big)
\end{split} 
\end{equation} 
with ${\bf x}= (x_0, \cdots, x_{t-2},x_t)$ and ${\bf y}= (y_0,
\cdots, y_{t-2},y_t)$ two sets of correlated normally distributed
variables which we choose to normalize. 
The correlations matrices between the different time steps are then
given by the elements ${\bf C}_{h}(t,t')=E[x_{t}x_{t'}]$ and ${\bf
C}_{\theta}(t,t')= E[y_{t}y_{t'}]$.

This concludes the explanation of the recursive scheme from which
the order parameters of the system can be obtained. A practical
difficulty that remains is the calculation of the explicit
correlations at different time steps. As an illustration we present
the first few time steps in Appendix A.

\subsection{Stationary equations}

Stationary equations for non-zero temperature can, in principle,
be obtained from the zero temperature expressions obtained in
\cite{jo} by introducing auxiliary thermal fields and averaging
over them. However, 
extreme care has to be taken concerning these thermal
fluctuations. The point is that in the case of the stationary limit
at zero temperature it is well known that there is no difference
beween the neural activity and the Edwards-Anderson order parameter,
but for non-zero temperatures they become distinct. Therefore,
in order to obtain the stationary limit, we assume that for
$t\rightarrow\infty$ the local fields can be replaced by their
thermal averages 
\begin{equation} 
\lim_{t\rightarrow\infty}[\sigma^{n}_{i}(t+1)]_{\Psi} 
=\lim_{t\rightarrow\infty}[g_{\Psi}^{n}\left(h_{i}(t),
\theta_{i}(t)\right)]_{\Psi}
\approx \lim_{t\rightarrow\infty}[g_{\Psi}^{n}\left([
h_{i}(t)]_{\Psi},[\theta_{i}(t)]_{\Psi}\right)]_{\Psi} 
\quad n=1,2\,. 
\end{equation} 

The first step is to find the thermally averaged fields. Starting
from the residual overlaps (\ref{rs}) we write (we focus on the
$h$-quantities) 
\begin{equation} 
\label{recres} 
[r^{\mu}(t+1)]_{\Psi} 
      =[\tilde{r}^{\mu}(t)]_{\Psi}+[\chi_{h}(t)r^{\mu}(t)]_{\Psi} 
\end{equation} 
where now 
\begin{equation} 
[\tilde{r}^{\mu}(t)]_{\psi}= 
\lim_{N\rightarrow\infty}\frac{1}{a^{2}\sqrt{N}} 
\sum_{i}\xi_{i}^{\mu} 
[g_{\Psi}(\tilde{h}_{N,i}^{\mu}(t),\tilde{\theta}_{N,i} 
				     ^{\mu}(t))]_{\Psi} 
       \stackrel{\mathcal Pr}{=} 
	      \mathcal{N}\left(0,\frac{1}{a^3}q_{1}(t+1)\right) 
\end{equation} 
with 
\begin{equation} 
q_{1}(t)=\left\langle\!\left\langle\,
[\sigma_{i}(t)]_\Psi^2\,\right\rangle\!\right\rangle\
\,. 
\end{equation} 
The recursion relation for the field then becomes 
\begin{equation} 
[h_{i}(t+1)]_{\Psi} 
=\frac{\xi_{i}}{a}m(t+1)+\chi_{h}(t)\left[h_{i}(t)
-\frac{\xi_{i}}{a}m(t)+\frac{\alpha}{a}\sigma_{i}(t)\right]_{\Psi}
+\mathcal{N}\left(0,\frac{q_{1}(t+1)}{a}\right) 
\end{equation} 
where we have used the LLN for $m(t)$ and $\chi_{h}(t)$, viz. 
\begin{equation} 
\begin{split} 
  m(t) &= \frac{1}{aN} \sum_i \xi_{i} \sigma_i(t) 
      \stackrel{\mathcal Pr}{=} \frac{1}{a}\left\langle\!\left\langle
      \xi_{i}\,[\sigma_i(t)]_{\Psi}\,\right\rangle\!\right\rangle\\ 
  \chi_h(t) &= \frac{1}{a^2N} \sum_i (\xi^\mu_i)^2 
		   \frac{\partial g_\Psi(t)}{\partial h_i(t)} 
     \stackrel{\mathcal Pr}{=}
     \frac{1}{a}\left\langle\!\left\langle\,\left[ 
	\frac{\partial g_\Psi(...)}{\partial h}\right]_{\Psi}\, 
	\right\rangle\!\right\rangle \,. 
\end{split} 
\end{equation} 
Completely analogous equations can be written down for
$\izq\theta_{i}(t+1)\der$ and $\izq\tilde{s}^{\mu}(t+1)\der$ with 
\begin{equation} 
p_{1}(t)=\left\langle\!\left\langle\,[\sigma_{i}^2(t)]_\Psi^2\, 
      \right\rangle\!\right\rangle\ \,. 
\end{equation} 

Next, in order to obtain the stationary limit we assume that the
thermal averages of the local fields and of the order parameters
do not change in time. This leads to 
\begin{equation} 
\begin{split} 
[h]_{\Psi}& 
    =\frac{\xi}{a}m+\mathcal{N}\left(0,\frac{\alpha
    q_1}{a^2(1-\chi_{h})^2}\right)+\frac{\alpha}{a}\eta_{h}[\sigma]_{\Psi}
   \equiv \tilde{h}+\frac{\alpha}{a}\eta_{h}[\sigma]_{\Psi} 
   \\ 
[\theta]_{\Psi}& 
   =\eta l+\mathcal{N}\left(0,\frac{\alpha
   p_1}{\tilde{a}^2(1-\chi_{\theta})^2}\right)+\frac{\alpha}
   {\tilde{a}}\eta_{\theta}[\sigma^2]_{\Psi}
\equiv
\tilde{\theta}+\frac{\alpha}{\tilde{a}}\eta_{\theta}[\sigma^2]_{\Psi}
\\ 
\end{split} 
\end{equation} 
with $\eta_{x}=\chi_x/(1-\chi_x)$, $x=h,\theta$. The solution for
$[\sigma]_{\Psi}$ is a self-consistent equation of the form 
\begin{equation} 
[\sigma]_{\Psi}= 
V_{\beta}\left(\tilde{h}+\frac{\alpha}{a}\eta_{h}[\sigma]_{\Psi}, 
	 \tilde{\theta} 
+\frac{\alpha}{\tilde{a}}\eta_{\theta}[\sigma^2]_{\Psi}\right)\,. 
\end{equation} 

In order to solve this equation we consider the case $T=0$. There
the averages over the thermal noise disappear, yielding 
\begin{equation} 
\lim_{\beta\rightarrow\infty}V_{\beta}\left(h,\theta\right) 
    =\mbox{sign}(h)\Theta(|h|+\theta) 
\end{equation} 
and a Maxwell construction \cite{SF93} leads to 
\begin{equation} 
\sigma=\mbox{sign}(h) \Theta(|h|+\theta+\Delta) 
\end{equation} 
where 
\begin{equation}
\Delta=\frac{1}{2}\left(\frac{\alpha}{a}\eta_{h}
+\frac{\alpha}{\tilde{a}}\eta_{\theta}\right)
\, . 
\end{equation} 
We introduce then a set of auxiliary thermal fields and average
over them to arrive at the fixed point equations 
in the limit $N \to \infty$ (dropping the index i) 
\begin{eqnarray} 
m&=&\left\langle\left\langle\frac{\xi}{a}\int Dz \int Dy \,
V_{\beta}(h(z),\theta(y)+\Delta)\right\rangle\right\rangle 
\label{fp1}\\ 
l&=&\left\langle\left\langle\eta\int Dz \int Dy \,
W_{\beta}(h(z),\theta(y)+\Delta)\right\rangle\right\rangle \\ 
q&=&\left\langle\left\langle\int Dz \int Dy \,
W_{\beta}(h(z),\theta(y)+\Delta)\right\rangle\right\rangle \\ 
\chi_{h}&=&\frac{1-\chi_h}{\sqrt{\alpha
q_1}}\left\langle\left\langle\int Dz \,z \int Dy \, 
V_{\beta}(h(z),\theta(y)+\Delta)\right\rangle\right\rangle \\ 
\chi_{\theta}&=&\frac{1-\chi_\theta}{\sqrt{\alpha p_1}} 
\left\langle\left\langle\int Dz \int Dy \, y \, 
W_{\beta}(h(z),\theta(y)+\Delta)\right\rangle\right\rangle 
\label{fp5} 
\end{eqnarray} 
with 
\begin{equation} 
h(z)= 
   \frac{\xi}{a}m + \frac{\sqrt{\alpha q_1}}{{a(1-\chi_{h})}} z 
\, ,\quad 
\theta(y)= 
      \eta l + 
  \frac{\sqrt{\alpha p_1}}{\tilde{a}(1-\chi_{\theta})} y \, . 
\end{equation} 
and $q_{1}$ and $p_1$ given by 
\begin{equation} 
\begin{split} 
q_{1}&=\left\langle\!\left\langle\int Dz \int Dy\,
V_{\beta}^{2}(h(z),\theta(y)+\Delta)\right\rangle\!\right\rangle\\ 
p_{1}&=\left\langle\!\left\langle\int Dz \int Dy\,
W_{\beta}^{2}(h(z),\theta(y)+\Delta)\right\rangle\!\right\rangle
\,. 
\end{split} 
\end{equation} 
Furthermore, in agreement with the fluctuation-dissipation theorem
\cite{FH} 
\begin{equation} 
\chi_{h}=\frac{\beta}{a}(q-q_{1})\qquad\chi_{\theta}
=\frac{\beta}{\tilde{a}}(q-p_{1})
\end{equation} 
with $\chi_h$ and $\chi_\theta$ the susceptibilities defined before,
in the stationary limit. 
It is interesting to remark that the equations
(\ref{fp1})-(\ref{fp5}) 
are the same as the fixed-point equations derived from a replica
symmetric thermodynamic approach with a hamiltonian for sequential
dynamics~\cite{BV03}.

\subsection{Numerical results and simulations} 
The results derived in Sections \ref{sec:mod}-\ref{sec:prob}
and apppendix A have been studied numerically and compared with
simulations for systems up to $N=7000$ neurons averaged over $500$
runs. 

As in the $T=0$ case the first few time steps, given by the explicit
formula in appendix A, and the simulations agree very well over
the whole range of $\alpha$. As an illustration we refer to Fig.~1
presenting the order parameters as a function of $\alpha$ 
for uniform patterns, $m_0=l_0=0.6, q_0=0.5$ and $T=0.5$. 
We remark that the critical capacity for this system is
$\alpha_{c}\simeq 0.06$ (\cite{BV03}). We then learn that the
first time steps do give us a reasonable estimate for the critical
capacity, especially through the order parameter $l$. 
\begin{figure}[ht] 
\centering\includegraphics[angle=270,width=.48\textwidth]{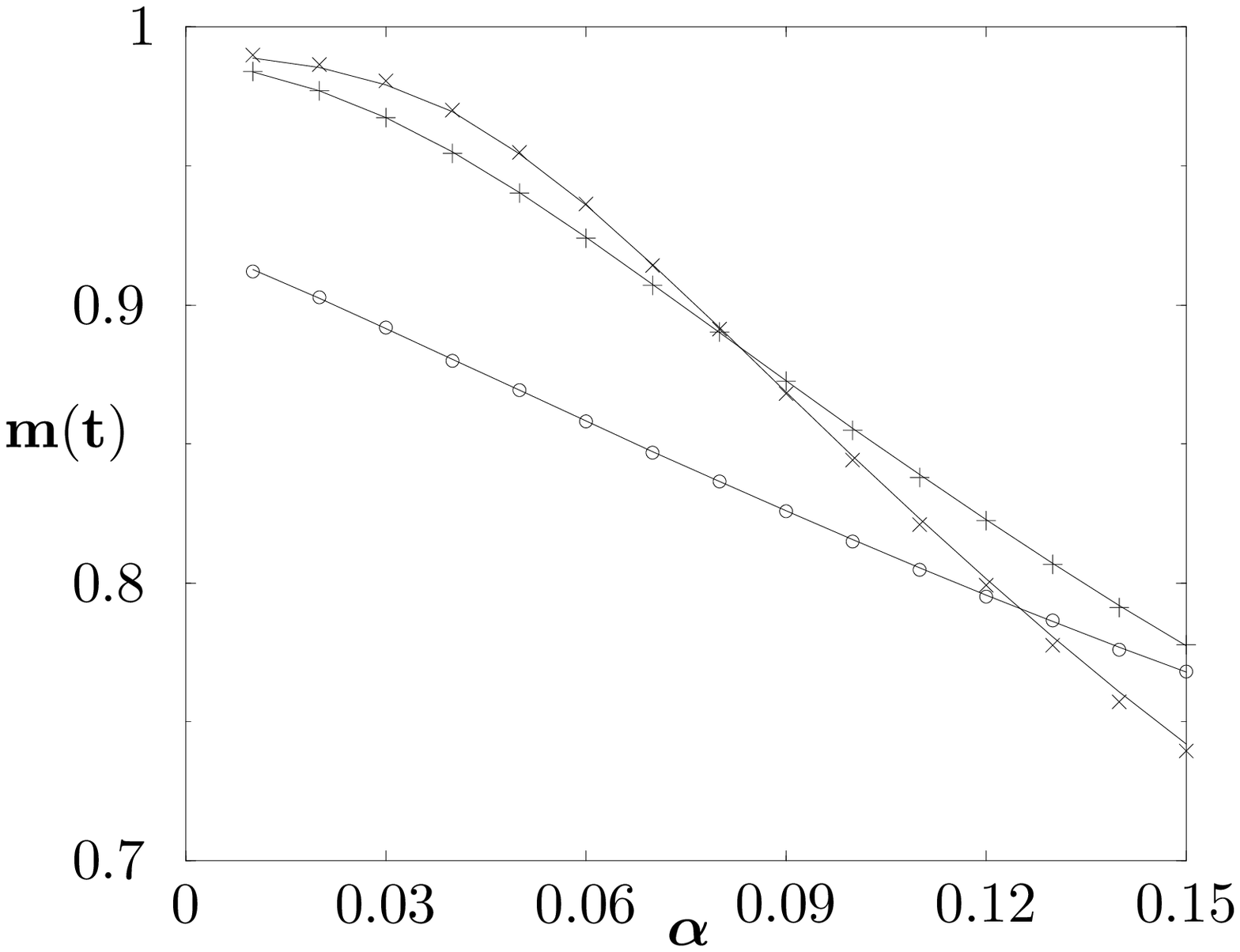} 
\hfill
\centering\includegraphics[angle=270,width=.48\textwidth]{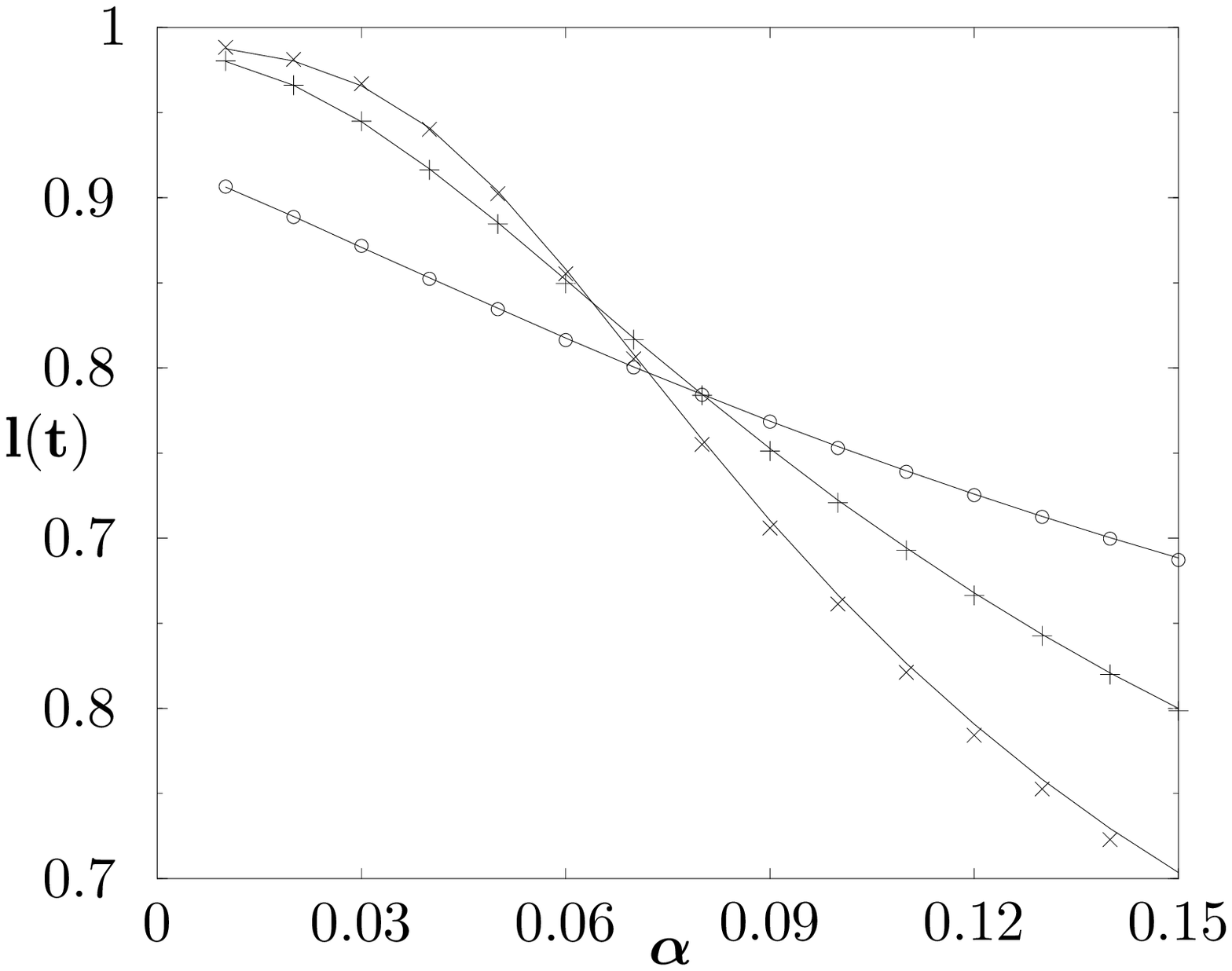}
\\
\centering\includegraphics[angle=270,width=.48\textwidth]{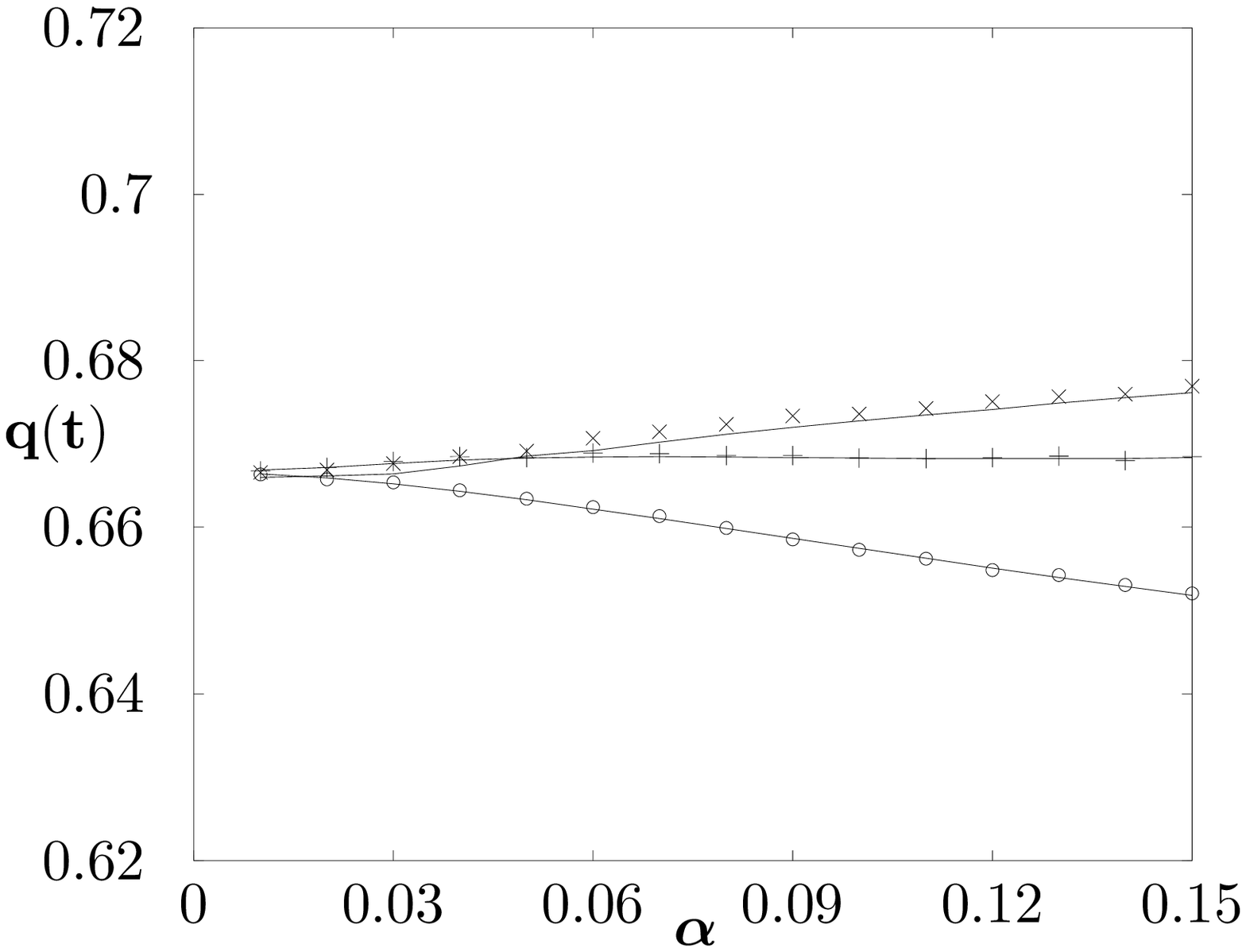} 
\caption{\footnotesize Order parameters $m(t)$, $l(t)$ and $q(t)$
as a function of the capacity $\alpha$ for the first three time
steps for uniform patterns ($a=2/3$), $m_0=l_0=0.6, q_0=0.5$
and $T=0.5$. Theoretical results (solid lines) versus simulations
(time $1,2$ and $3$ represented by a circle, a plus respectively
a times symbol) are shown.} 
\end{figure}

In Fig.~2 we examine the order parameters $m$ and $l$ in the
quadrupolar phase ($m=0,l>0$) versus the paramagnetic phase
($m=0,l=0$), for several values of $\alpha$. We see that a few
time steps do give us already the characteristic behaviour. When
time increases $m$ decreases while $l$ differentiates between
the phases, as is seen in the theoretical results as well as in
the simulations. For the quadrupolar phase ($\alpha=0.001$) $l$
increases, deep inside the paramagnetic phase ($\alpha=0.1$) $l$
decreases, while in the intermediate region 
($\alpha=0.01$) the rate of increase of $l$ quickly diminishes and
$l$ itself goes to zero. 
\begin{figure}[ht] 
\centering\includegraphics[angle=270,width=.6\textwidth]{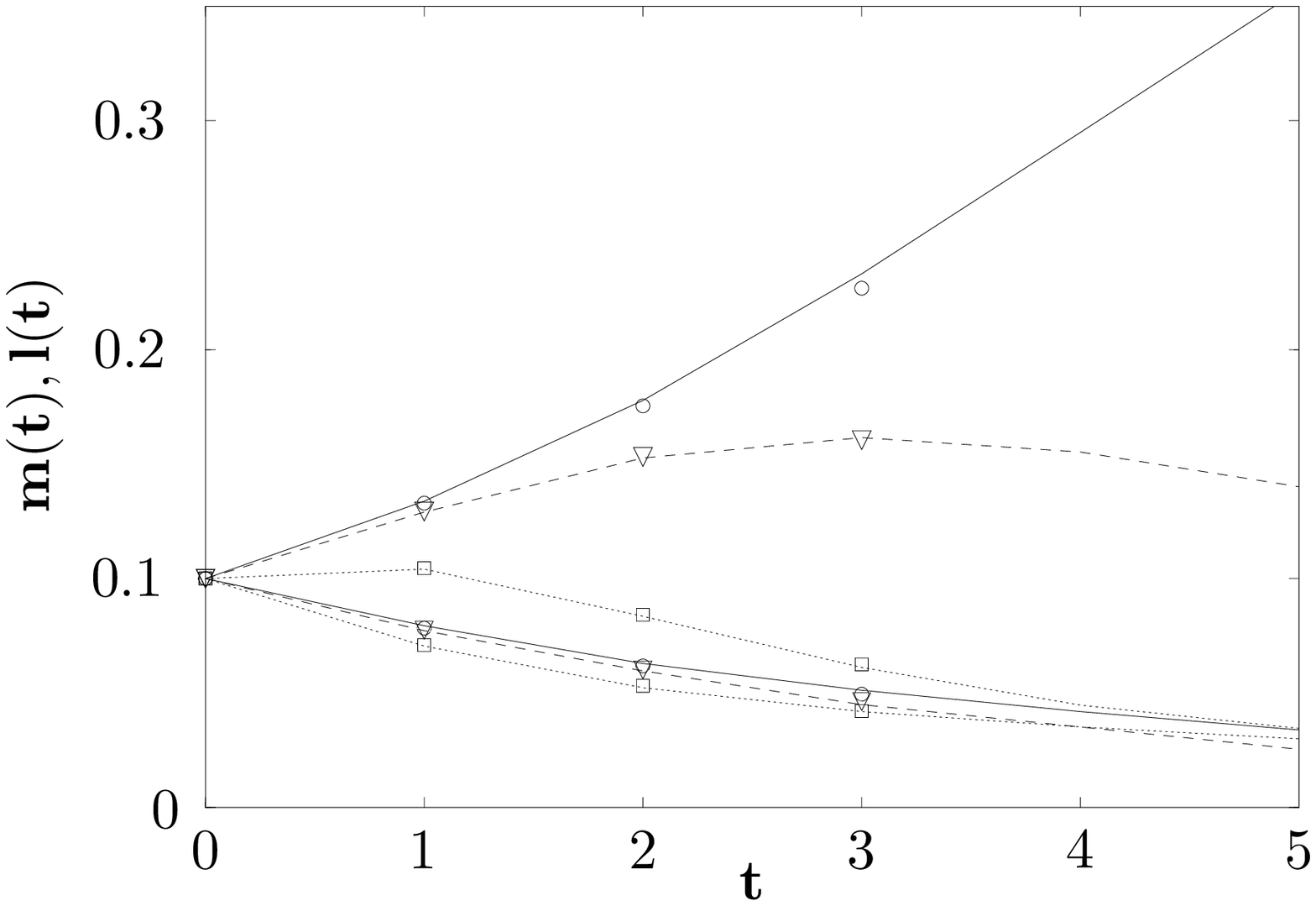} 
\caption{\footnotesize Order parameters $m(t)$ (bottom three lines)
and $l(t)$ (top three lines) as a function of time for uniform
patterns ($a=2/3$), $m_0=l_0=0.1, q_0=0.5$, $T=1.1$ and several
values of $\alpha$. Theoretical results (open symbols) are shown
versus simulations (full lines for $\alpha=0.001$, dashed lines
for $\alpha=0.01$ and dotted lines for $\alpha=0.1$).} 
\end{figure}

At this point, we remark that there is a small but visible
discrepancy between the theory and simulations especially in
$l(3)$. It is of the order $O(10^{-3})$ and attributed to finite-size
effects. In this respect we report that earlier calculations
\cite{BJSu} of further time steps of the overlap order parameter in
the Hopfield model also showed a discrepancy between simulations
and numerical results, e.g., for $\alpha=0.10$ the difference
$m_{th}(3)-m_{sim}(3)$ is again of the order $O(10^{-3})$. However,
for $0.1<m_0<0.4$, the difference $m_{th}(4)-m_{sim}(4)$ is about
$0.018$ and always positive both for $N=6000$ and $N=7000$. This,
and the fact that the SNA does not give a closed form solution of
the dynamics, has motivated us to look in Section~4 at a second,
alternative approach for solving the dynamics.

\section{The generating functional approach} 

\label{sec:genfunc} 
The second approach we use to discuss the dynamics of the BEG-model
is an extension of the generating functional approach (GFA)
\cite{SMR73} \cite{D78} (for a recent review see, e.g., \cite{C01}
and references therein). It is an exact procedure based on the
idea of looking at the probability for finding a microscopic path
in time 
\begin{equation} 
\mbox{Prob}[\bsigma_N(0),\ldots,\bsigma_N(t)]= 
 \mbox{Prob}(\bsigma(0)) \prod_{t'=1}^{t} \prod_{i=1}^N 
   \mbox{Prob} \left(\sigma_i(t') = s_k \in \mathcal{S}|
   \bsigma_N(t'-1) \right) 
\label{eq:transpath} 
\end{equation} 
where the transition probabilities are given by
(\ref{eq:trans})-(\ref{eq:energy}) with the local fields
$h_i(\bsigma_N(t))$ and $\theta_i(\bsigma_N(t))$ replaced by 
\begin{equation} 
h_{i}(\bsigma_N(t))=\sum_{j\neq
i}J_{ij}\sigma_{j}(t)+\gamma_{h,i}(t)\qquad \theta_{i}(t)=\sum_{j\neq
i}K_{ij}\sigma_{j}^{2}(t)+\gamma_{\theta,i}(t)\,. 
\end{equation} 
Here the $\gamma_{h,i}(t)$ and $\gamma_{\theta,i}(t)$ are
time-dependent external fields introduced in order to define a
response function. 

The basic tool is then the generating functional 
\begin{align} 
Z_N[\Psi, \Phi]
&= 
\left\langle
\exp\left\{-i\sum_{s=1}^{t}\sum_{i=1}^{N}\left(\psi_{i}(s)\sigma_{i}(s)
    +\phi_{i}(s)\sigma_{i}^{2}(s)\right)\right\} 
		   \right\rangle_{paths} \nonumber \\ 
&= 
 \sum_{\bsigma_N(0)} \ldots \sum_{\bsigma_N(t)} 
  \mbox{Prob}[\bsigma_N(0), \ldots,\bsigma_N(t)] 
  \nonumber \\
& \qquad \qquad
 \times \exp\left\{-i\sum_{s=1}^{t}\sum_{i=1}^{N}\left(\psi_{i}(s)\sigma_{i}(s)
		 +\phi_{i}(s)\sigma_{i}^{2}(s)\right)\right\}
   \label{genfun} 
\end{align} 
from which we can find all the relevant observables by calculating
derivatives with respect to $\{\psi_i(t)\}$ and $\{\phi_i(t)\}$ and
taking these zero afterwards. In the sequel we omit the subscript
indicating the path average. This path average of the spin, the
correlation functions and the response functions are given by
\begin{equation} 
\begin{matrix}
\displaystyle
\left \langle \sigma_i(t)\right \rangle 
    =i \lim_{\Psi,\Phi\rightarrow 0} 
  \frac{\partial Z_N[\Psi, \Phi]}{\partial \psi_{i}(t)} \, , 
&
\displaystyle
  \left \langle \sigma_i^2(t)\right \rangle= 
   i \lim_{\Psi,\Phi\rightarrow 0} 
	\frac{\partial Z_N[\Psi, \Phi]}{\partial \phi_{i}(t)} \, , 
\\ 
\displaystyle
C_{h,ij}(t,t') 
   =-\lim_{\Psi,\Phi\rightarrow 0} 
\frac{\partial^{2} Z_N[\Psi,\Phi]} 
		{\partial\psi_{i}(t)\partial\psi_{j}(t')}\, , 
& 
\displaystyle
C_{\theta,ij}(t,t')= 
   -\lim_{\Psi,\Phi\rightarrow 0} 
   \frac{\partial^{2} Z_N[\Psi, \Phi]} 
   {\partial\phi_{i}(t)\partial\phi_{j}(t')} \, , 
\\ 
\displaystyle
G_{h,ij}(t,t') 
    =i\lim_{\Psi,\Phi\rightarrow 0} 
    \frac{\partial^{2} Z_N[\Psi, \Phi]} 
{\partial \psi_{i}(t)\partial\gamma_{h,j}(t')}\, , 
& 
\displaystyle
G_{\theta,ij}(t,t')= 
    i\lim_{\Psi,\Phi\rightarrow 0} 
	\frac{\partial^{2}Z_N[\Psi, \Phi]} 
	    {\partial \phi_{i}(t)\partial\gamma_{\theta,j}(t')}\,. 
	    \label{cordef3} 
\end{matrix}
\end{equation} 

The generating function is averaged over all pattern realizations,
i.e. over the disorder, before the path average is taken. In the
limit $N \to \infty$, this results in a theory with effective single
spin local fields which, for the model at hand, are given by 
\begin{align} 
h(t) &=\frac{\xi}{a}m(t)+\gamma_h(t)+ 
      \frac{\alpha}{a}\sum_{s'=0}^{t}R_h(t,s')\sigma(s') 
		    +\sqrt{\alpha}\,z_{h}(t) 
   \label{hpath} \\ 
\theta(t)&=\eta l(t) + \gamma_{\theta}(t)+ 
   \frac{\alpha}{\tilde{a}}\sum_{s'=0}^{t}R_{\theta}(t,s')\sigma^{2}(s')
	    +\sqrt{\alpha}\,z_{\theta}(t) 
   \label{thetapath} 
\end{align} 
where the noises $z_{h}(t),z_{\theta}(t)$ are a set of correlated
normally distributed variables with the measure 
\begin{align} 
w({\bf z}_h,{\bf z}_{\theta})&= 
\frac{\exp\left\{-\frac{1}{2}\sum_{s,s'=0}^{t-1} 
	   z_{h}(s)[{\bf w}_{h}^{-1}](s,s')z_{h}(s') 
     -\frac{1}{2}\sum_{s,s'=0}^{t-1} 
z_{\theta}(s)[{\bf w}_{\theta}^{-1}](s,s')z_{\theta}(s') \right\} } 
  {\sqrt{(2\pi)^{t}\det({\bf w}_{h})}\, 
      \sqrt{(2\pi)^{t} \det({\bf w}_{\theta})}}\\ 
{\bf w}_{h}&= 
     \left(a{\bf I}-{\bf G}_{h}\right)^{-1} {\bf C}_{h} 
       \left(a{\bf I}-{\bf G}_{h}^{\dagger}\right)^{-1}\\ 
{\bf w}_{\theta}&= 
 \left(\tilde{a}{\bf I}-{\bf G}_{\theta}\right)^{-1}{\bf C}_{\theta}
	\left(\tilde{a}{\bf I}-{\bf
	G}_{\theta}^{\dagger}\right)^{-1}\,. 
\end{align} 
The retarded self-interactions appearing in
(\ref{hpath})-(\ref{thetapath}) are then given by 
\begin{equation} 
{\bf R}_h=\left(a{\bf I}-{\bf G}_{h}\right)^{-1}{\bf G}_{h}\, , 
\qquad 
{\bf R}_{\theta}= 
\left(\tilde{a}{\bf I}-{\bf G}_{\theta}\right)^{-1}{\bf G}_{\theta} 
     \,. 
\end{equation} 
Furthermore, the order parameters can be written as 
\begin{eqnarray} 
 m(t)&=&i \lim _{N \to \infty}\lim_{\Psi,\Phi\rightarrow 0} 
 \frac{1}{aN} \sum_i\xi_i 
      \frac{\partial {\overline {Z_N[\Psi, \Phi]}}}{\partial
      \psi_{i}(t)} 
 =\frac{1}{a}\left \langle \! \left \langle \xi \sigma(t) 
	       \right \rangle \! \right \rangle 
	 \\ 
 l(t) &=&i \lim _{N \to \infty}\lim_{\Psi,\Phi\rightarrow 0} 
  \frac{1}{N} \sum_i\eta_i 
  \frac{\partial \overline{Z_N[\Psi, \Phi]}}{\partial \phi_{i}(t)} 
  =\left \langle \! \left \langle \eta \sigma^2(t) 
       \right \rangle \! \right \rangle 
	   \\ 
C_{h}(t,t')&=&-\lim _{N \to \infty}\lim_{\Psi,\Phi\rightarrow 0} 
    \frac{1}{N}\sum_i 
\frac{\partial^{2} \overline{Z_N[\Psi,\Phi]}} 
		  {\partial\psi_{i}(t)\partial\psi_{i}(t')} 
=\left \langle \! \left \langle \sigma(t) \sigma(t') 
	\right \rangle \! \right \rangle 
		       \\ 
C_{\theta}(t,t')&=&-\lim _{N \to \infty}\lim_{\Psi,\Phi\rightarrow
0} 
\frac{1}{N}\sum_i\frac{\partial^{2}\overline{ Z_N[\Psi, \Phi]}} 
		    {\partial \phi_{i}(t)\partial\phi_{i}(t')} 
=\left \langle \! \left \langle \sigma^2(t) \sigma^2(t') 
    \right \rangle \! \right \rangle 
		  \\ 
G_{h}(t,t')&=&i\lim _{N \to \infty}\lim_{\Psi,\Phi\rightarrow 0} 
\frac{1}{N}\sum_i\frac{\partial^{2}\overline{ Z_N[\Psi, \Phi]}} 
       {\partial \psi_{i}(t)\partial\gamma_{h,i}(t')} 
=\left \langle \! \left \langle 
	  \frac{\partial}{\partial\gamma_{h}} \sigma(t') 
	   \right \rangle \! \right \rangle 
	    \\ 
G_{\theta}(t,t')&=&i\lim _{N \to\infty}\lim_{\Psi,\Phi\rightarrow
0} 
  \frac{1}{N}\sum_i\frac{\partial^{2}\overline{Z_N[\Psi, \Phi]}} 
	 {\partial \phi_{i}(t)\partial\gamma_{\theta,i}(t')} 
=\left \langle \! \left \langle 
   \frac{\partial}{\partial\gamma_{\theta}} \sigma^2(t') 
		\right \rangle \! \right \rangle 
\end{eqnarray} 
where the overline denotes the disorder average and $\left
\langle\!\left \langle \cdot \right \rangle\!\right \rangle $
denotes 
the averages over the noises ${\bf z}_h,{\bf z}_{\theta}$, and
the initial conditions and the condensed pattern, according to the
measure 
\begin{equation} 
\label{cata} 
\left \langle\!\left \langle f[\sigma(0),\ldots,\sigma(t)]\right
\rangle\!\right \rangle= 
 \left\langle\int d{\bf z}_{h} d{\bf z}_{\theta} w[{\bf z}_{h},{\bf
 z}_{\theta}] 
 \mbox{Prob}[\sigma(0),\ldots,\sigma(t)|{\bf z}_{h},{\bf z}_{\theta}]
\,\,f[\sigma(0),\ldots,\sigma(t)]\right\rangle_{\xi} 
\end{equation} 
Here 
\begin{multline} 
\label{cata2} 
\mbox{Prob}[\sigma(0),\ldots,\sigma(t)|{\bf z}_{h},{\bf z}_{\theta}]
\\
=\sum_{\sigma(0),\ldots,\sigma(t)}\,\, 
\mbox{Prob}(\sigma(0))\prod_{s=0}^{t-1}\frac{\exp\left(\beta\sigma(s+1)h(s)
				 +\beta\sigma^{2}(s+1)\theta(s)\right)}
	{2\exp\left(\beta\theta(s)\right)\cosh(\beta h(s))+1} 
\end{multline} 
with $h(s)$ and $\theta(s)$ given by (\ref{hpath}) and
(\ref{thetapath}), $\gamma_h, \gamma_{\theta}=0$. For more details
of this calculation we refer to Appendix B. 

For later convenience we remark that $G_{h}(t,t')=0$ for $t \leq
t'$ due to causality (analogously for $G_{\theta}(t,t')$), and that
for $t>t'$ 
\begin{eqnarray} 
G_{h}(t,t')&=& 
\beta\left\langle\!\left\langle 
\sigma(t)\left[\sigma(t'+1)- V_{\beta}(h(t'),
\theta(t'))\right]\right\rangle\!\right\rangle 
\label{resph} \\ 
G_{\theta}(t,t')&=& 
\beta\left\langle\!\left\langle\ 
\sigma^{2}(t)\left[\sigma^{2}(t'+1)-W_{\beta}(h(t'), \theta(t')) 
\right]\right\rangle\!\right\rangle \,. 
\end{eqnarray} 
This set of equations completely solves the dynamics for the
BEG-model. 
We calculate the first few time steps in Appendix C, in order to
compare the results with those of the SNA method.

\section{Comparison between SNA and GFA} 

\begin{figure}[p!] 
\centering\includegraphics[angle=270,width=.45\textwidth] 
{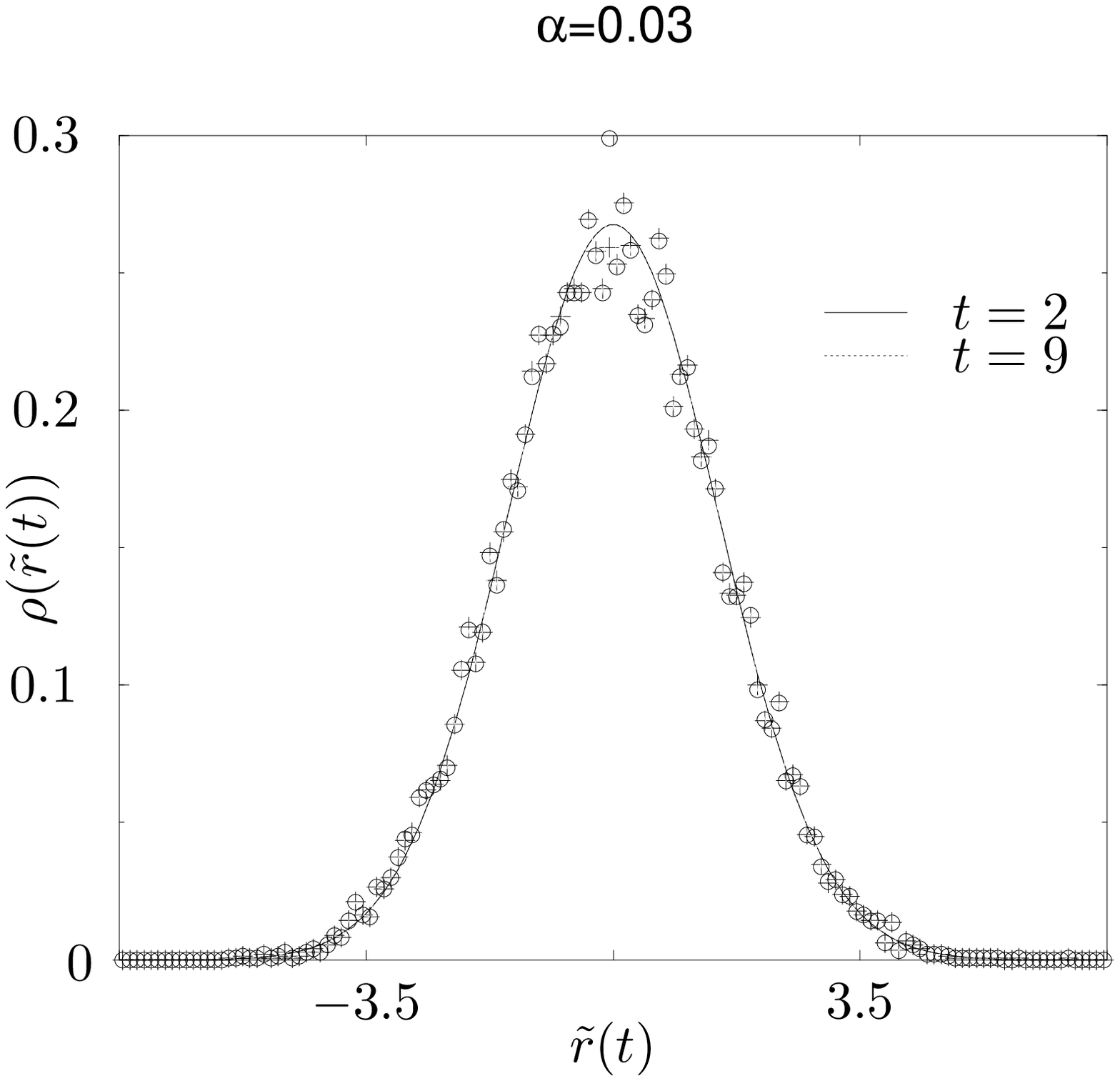}
\hfill
\includegraphics[angle=270,width=.45\textwidth] 
{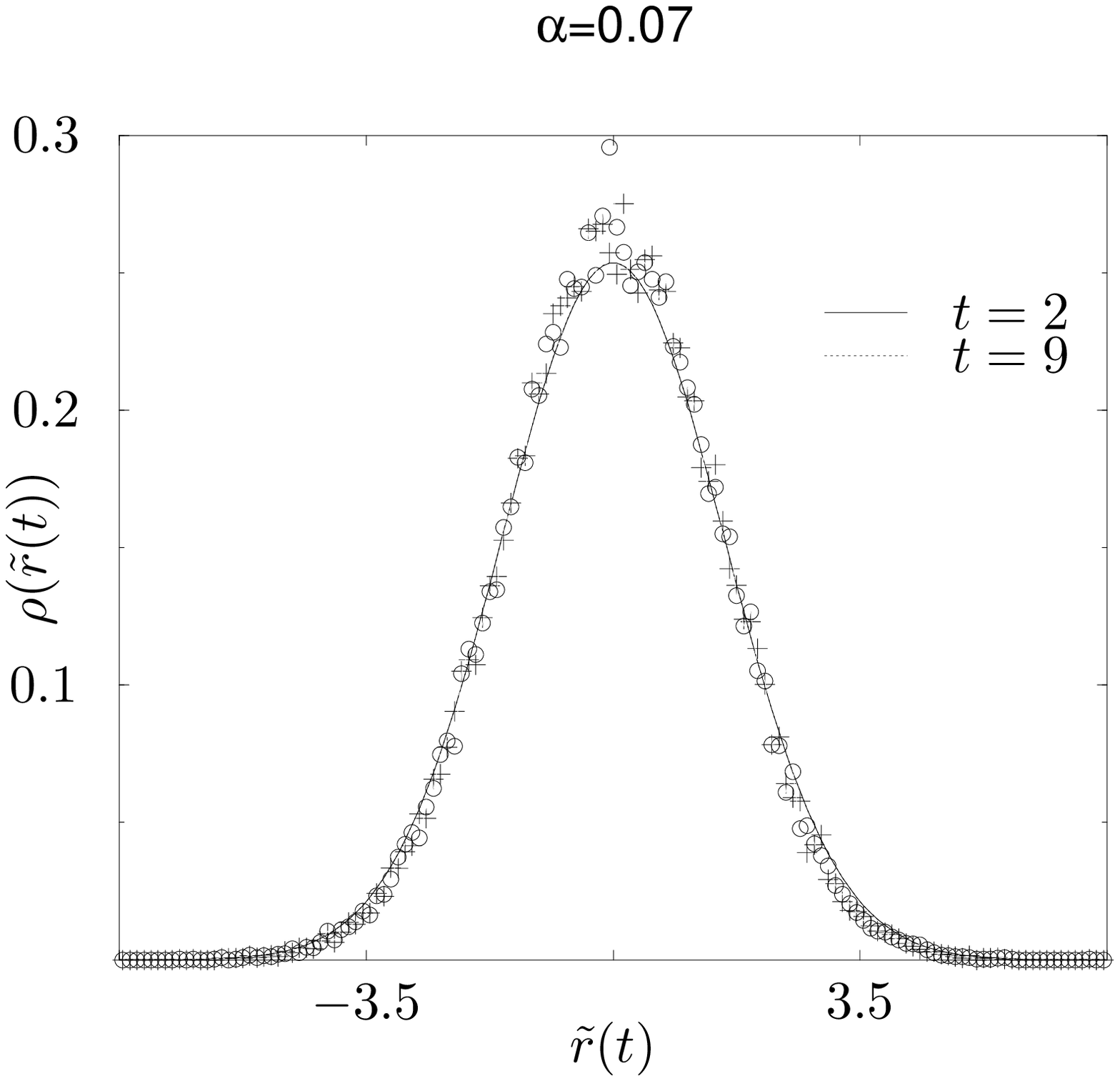}\\ 
\vspace{.5cm}
\includegraphics[angle=270,width=.45\textwidth] 
{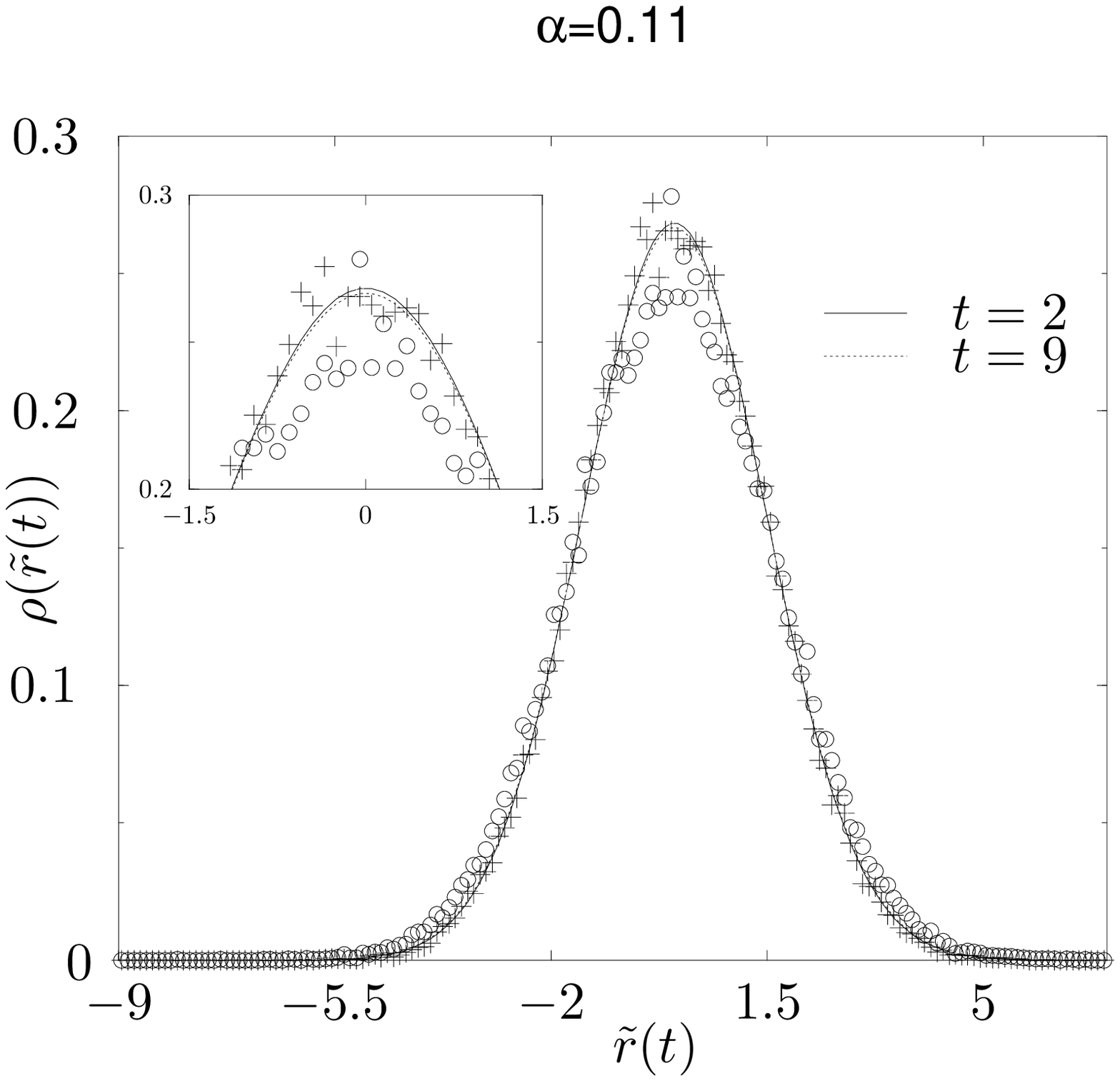} 
\hfill
\includegraphics[angle=270,width=.45\textwidth] 
{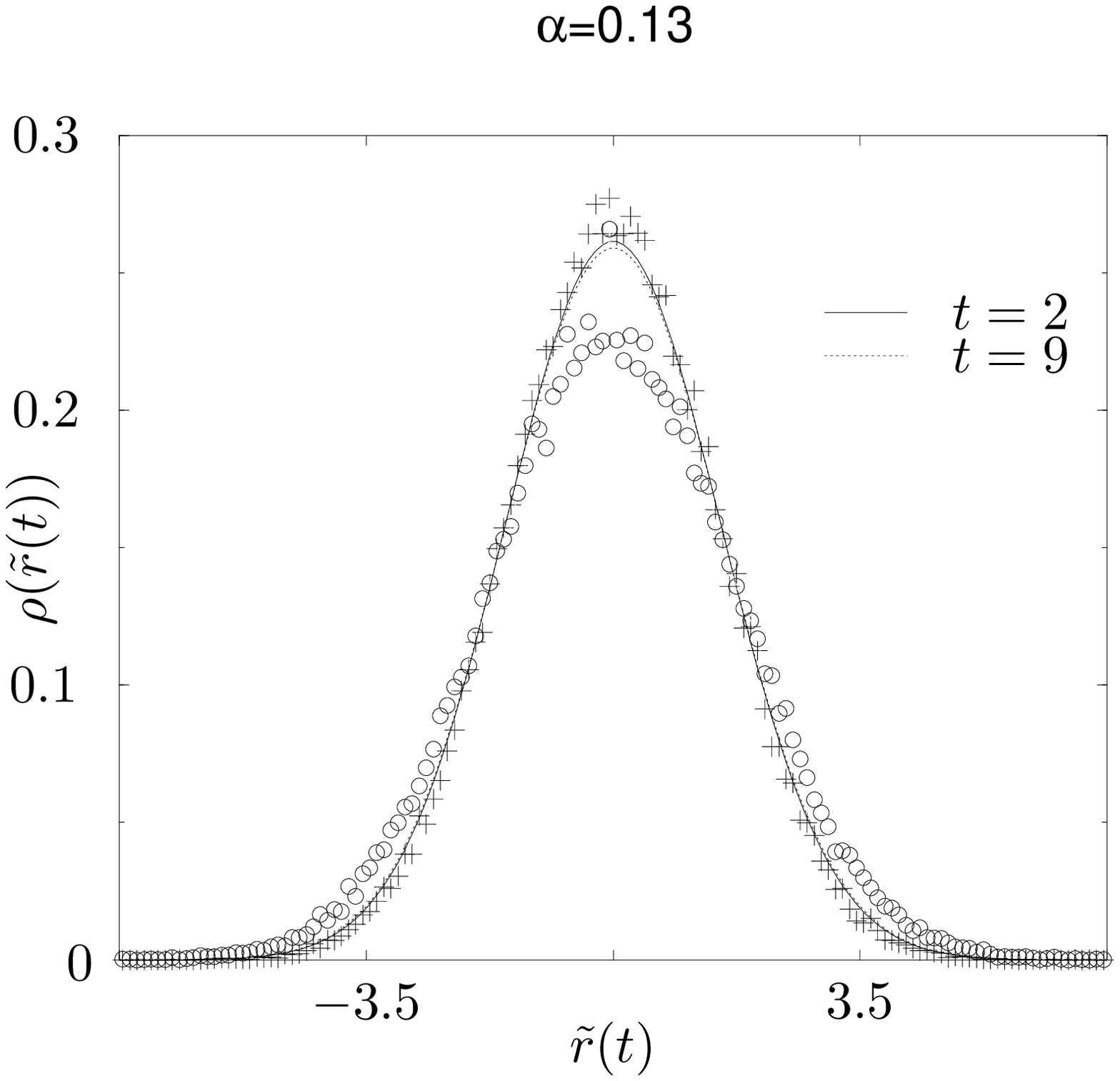} 
\caption{\footnotesize Simulations of the distribution of the
residual overlap $\tilde{r}(t)$ for time steps $2$ and $9$, compared
with the limiting normal distribution (full and dotted lines)
as a function of the capacity $\alpha$. The simulations 
are done for $N=2000$ neurons averaged over 250 samples. The initial
conditions for the dynamics 
are $m_{0}=l_{0}=0.6, q_{0}=0.5, a=2/3$ and $T=0.2$.} 
\end{figure} 

By comparing the explicit calculations described in the Appendices
A and C for the first two time steps in the dynamics of the BEG
model we find that they are identical and that in the GFA response
functions, restricting ourselves for the moment to the $h$-field, 
\begin{equation} 
G_{h}(1,0)={a}\chi_h(0) \qquad G_{h}(2,1)={a}\chi_h(1)\,. 
\end{equation} 
The third time step requires a bit more work. Looking at the response
function in general (recall (\ref{resph})) and considering $t'=t-2$,
the sums over $\sigma(t)$ and $\sigma(t-1)$ can be done explicitly
because of the specific time behaviour of the local fields, leading
to 
\begin{equation} 
G_{h}(t,t-2)=0 \,. 
\end{equation} 
Using this, together with the causality requirement on $G_{h}$ and noting that
a similar treatment is valid for $G_\theta$,
we find that the result for the third time step in the GFA method
coincides with the corresponding SNA result. 
Comparing the next time steps explicitly is more
cumbersome. Also from the local fields as expressed by
Eqs.(\ref{hprob})-(\ref{thetaprob}) in the SNA approach and
Eqs.(\ref{hpath})-(\ref{thetapath}) in the GFA method no immediate
obvious analytic connection is apparent. Therefore, a separate
detailed analytic study of these relationships for the simpler
Hopfield model is appropriate \cite{BBVpre}. 

In the sequel we present some strong numerical evidence that the
SNA and the GFA approach will lead to slightly different results
for certain values of the model parameters, specifying spin-glass
behaviour, from the fourth time step onwards. This is consistent with
our remark at the end of Section 3 that a difference, although a very
small one, has been noticed between the SNA approach and simulations
at zero temperature in the fourth time step of the parallel Hopfield
dynamics \cite{BJSu}. This result was not entirely conclusive
because of the possible entanglement with finite-size effects.

Reviewing the SNA approach one knows that one of the basic
ingredients is the introduction of a modified residual overlap
$\tilde{r}(t)$ which converges to a normal distribution in the
thermodynamic limit (recall eqs.~(\ref{tilder})-(\ref{tildes})). 

We have done some numerical experiments for different values of
the model parameters comparing this limiting normal distribution
in the SNA method with simulations for different time steps. Some
results are shown in Fig.~3. We compare the normal distribution
with simulations for time steps $t=2$ and $t=9$ for systems with
$N=2000$ neurons averaged over $250$ runs for the initial conditions
$m_0=l_0=0.6, q_0=0.5, a=2/3$ and temperature $T=0.2$ as a function
of $\alpha$. We note that the critical $\alpha$ for these parameter
values is $\alpha_c =0.086$ \cite{BV03}.

We conclude that in the retrieval region ($\alpha_c < 0.086$)
the simulations results coincide quite well with the limiting
distribution, while in the spin-glass region, certainly from
$\alpha \sim 0.11$ onwards, the results for $t=9$ start diverting
systematically. We remark that a set of analogous results can be
obtained for $\tilde{s}(t)$. 
This may indicate that for $t>3$ the SNA works quite well in a
large part of the retrieval phase since most of the time the system
reaches the attractor already after a few time steps. However,
in the spin-glass phase the SNA method does not estimate correctly
the noise of the non-condensed patterns represented by $\tilde{r}(t)$,
suggesting that the assumption on the residual noise discussed above
is no longer fulfilled. Resolving these issues analytically is not
immediately straightforward and beyond the scope of the present
work. They will be worked out first for the 
Hopfield model~\cite{BBVpre}.

\section{Concluding remarks} 
The parallel dynamics of the fully connected BEG neural network
model for finite temperatures has been solved using both the signal
to noise analysis and the generating functional approach. 
It is shown that both methods give identical results up to the
first three time steps.  

Furthermore, by studying the distribution of the residual noise
in the signal to noise analysis numerically and comparing it with
further time steps up to $t=9$ it is found that the signal to noise
method as it is used in the literature does not estimate correctly
the modified residual overlaps in the spin-glass region. 
However, in a large part of the retrieval region it can be considered
as accurate. Numerical simulations confirm these findings. 
The signal to noise approach can be made exact by carefully studying
the long time correlations \cite{BBVpre}. 

\section*{Acknowledgments} 

The authors thank A.C.C. Coolen for useful discussions. This work
has been supported in part by the Fund of Scientific Research,
Flanders-Belgium. 

%%%%%%%%%%%%%%%%%%%%%%%%%%%%%%%%%%%%%%%%%%%%%%%%%%%%%%%%%%%%%%%%%%%%%

\appendix

\section{The first three time steps in the SNA} 

Employing the general recursive scheme of Section 3 evolution
equations are derived for the first three time steps of the dynamics
in the SNA approach. The following initial conditions are taken 
\begin{eqnarray} 
m^{1}(0)&=&m_{0},\quad l^{1}(0)=l_{0},\quad q(0)=q_{0}, 
	  \\ 
h_{i}(0)&=&\frac{\xi_i^1}{a}m_{0}+ 
		      \mathcal{N}\left(0,\frac{\alpha
		      q_{0}}{a^{2}}\right), 
     \quad 
\theta_{i}(0)=\eta_{i}^{1}l_{0}+ 
	    \mathcal{N}\left(0,\frac{\alpha
	    q_{0}}{\tilde{a}^{2}}\right), 
	   \\ 
D(0)&=&\frac{q_{0}}{a^{3}},\quad E(0)=\frac{q_{0}}{\tilde{a}}. 
\end{eqnarray} 
From now we forget about the superscript $1$ to indicate the
condensed pattern. Furthermore, we do not write down all details
but we refer to \cite{jo} for the analogous $T=0$ case. 

\subsection{First time step} 

From (\ref{defmt})-(\ref{deflt}) we immediately find 
\begin{eqnarray} 
m(1)&=&\frac{1}{a}\left\langle\!\left\langle 
    \xi \int Dz\int Dy \,\,
    V_{\beta}\big(h_{0}'(z),\theta_{0}'(y)\big) 
			   \right\rangle\!\right\rangle , 
	     \\ 
q(1)&=&\left\langle\!\left\langle 
     \int Dz\int Dy \,\, W_{\beta}\big(h_{0}'(z), \theta_{0}'(y)\big)
			       \right\rangle\!\right\rangle , 
			    \\ 
l(1)&=&\left\langle\!\left\langle 
    \eta\int Dz\int Dy
    \,\,W_{\beta}\big(h_{0}'(z),\theta_{0}'(y)\big) 
	       \right\rangle\!\right\rangle , 
\end{eqnarray} 
where 
\begin{equation} 
h_{0}'(z)=\frac{1}{a}\xi m_{0}+\sqrt{\alpha a D(0)}\, z, 
   \quad 
\theta_{0}'(y)=\eta l_{0}+\sqrt{\frac{\alpha}{\tilde{a}}E(0)}\, y 
\end{equation} 
and where $Dx={dx}\exp(-x^{2}/2)/\sqrt{2\pi}$ is the Gaussian
measure. 
Because of the modification of the local fields we know that 
\begin{eqnarray} 
\mbox{Cov}\big[\tilde{r}^{\mu}(0),r^{\mu}(0)\big] 
	 =E\big[\tilde{r}^{\mu}(0)r^{\mu}(0)\big] 
&=&E\big[\frac{1}{a^3}\sigma_{i}(0) 
      g_{\Psi}(\tilde{h}_{i}^{\mu}(0),\tilde{\theta}_{i}^{\mu}(0))
      \big]\\ 
\mbox{Cov}\big[\tilde{s}^{\mu}(0),s^{\mu}(0)\big] 
	 =E\big[\tilde{s}^{\mu}(0)s^{\mu}(0)\big] 
&=&E\big[\frac{1}{\tilde{a}}\sigma_{i}^2(0) 
      g_{\Psi}^2(\tilde{h}_{i}^{\mu}(0),\tilde{\theta}_{i}^{\mu}(0))
      \big] 
\end{eqnarray} 
Using the recursion relations (\ref{drec})-(\ref{erec}) we obtain 
\begin{eqnarray} 
D(1)&=&\frac{q(1)}{a^{3}}+\chi_{h}^{2}(0)D(0)+2\chi_{h}(0)R(1,0) 
       \\ 
E(1)&=&\frac{q(1)}{\tilde{a}}+\chi_{\theta}^{2}(0)E(0)+ 
				       2\chi_{\theta}(0)S(1,0)\, , 
\end{eqnarray} 
where the correlation parameters $R(t,t')$ and $S(t,t')$ are defined
by 
\begin{eqnarray} 
R(t,t')&=&\frac{1}{a^{3}}E\big[g_{\Psi}
(\tilde{h}(t-1),\tilde{\theta}(t-1))
g_{\Psi}(\tilde{h}(t'-1),\tilde{\theta}(t'-1))\big]
\\ 
R(t,0)&=&\frac{1}{a^{3}}
E\big[g_{\Psi}(\tilde{h}(t-1),
\tilde{\theta}(t-1))\sigma(0)\big]
\\ 
S(t,t')&=&\frac{1}{\tilde{a}}
E\big[g_{\Psi}^{2}(\tilde{h}(t-1),\tilde{\theta}(t-1))
g_{\Psi}^{2}(\tilde{h}(t'-1),\tilde{\theta}(t'-1))\big]
\\ 
S(t,0)&=&\frac{1}{\tilde{a}}
E\big[g_{\Psi}^{2}(\tilde{h}(t-1),
\tilde{\theta}(t-1))\sigma^{2}(0)\big]
\end{eqnarray} 
Since the density distribution of $h$ and $\tilde h$ are the same
in the limit $N \to \infty$ we have for the first time step 
\begin{equation} 
R(1,0)=\frac{1}{a^{3}}\left\langle\!\left\langle 
   \sigma(0)\int Dz\int Dy \,
   V_{\beta}\big(h_{0}'(z),\theta_{0}'(y)\big) 
		  \right\rangle\!\right\rangle 
\end{equation} 
\begin{equation} 
S(1,0)=\frac{1}{\tilde{a}}\left\langle\!\left\langle 
	\sigma^{2}(0)\int Dz\int Dy \, 
	   W_{\beta}\big(h_{0}'(z),\theta_{0}'(y)\big) 
		 \right\rangle\!\right\rangle 
\end{equation} 
Furthermore, since there are no feedback correlations at time zero 
\begin{eqnarray} 
\chi_{h}(0)&=&\frac{1}{a}\left\langle\!\left\langle 
     \int Dz\int Dy \, 
     \left.\Big(\frac{ 
     {\partial V_{\beta}(h(t),\theta(t))}} 
     {{\partial h(t)}}\Big)\right|_{(h'_0(z),\theta_0'(z))} 
	  \right\rangle\!\right\rangle 
\\ 
\chi_{\theta}(0)&=&\frac{1}{\tilde{a}}\left\langle\!\left\langle 
   \int Dz\int Dy 
  \left.\Big(\frac{ 
     {\partial W_{\beta}(h(t),\theta(t))}} 
     {{\partial \theta(t)}}\Big)\right|_{(h'_0(z),\theta_0'(z))} 
	     \right\rangle\!\right\rangle 
\end{eqnarray} 

\subsection{Second time step} 

First, we need the distribution of the local fields at time step
$1$ 
using (\ref{rec1})-(\ref{rec4}) 
\begin{eqnarray} 
h_{i}(1)&=& 
\frac{\xi_{i}}{a}m(1)+\frac{\alpha}{a}\chi_{h}(0)\sigma_{i}(0) 
	      +\mathcal{N}(0,\alpha a D(1)) 
 \\ 
\theta_{i}(1)&=&\eta_{i}l(1) 
+\frac{\alpha}{\tilde{a}}\chi_{\theta}(0)\sigma_{i}^{2}(0) 
	     +\mathcal{N}(0,\frac{\alpha}{\tilde a}E(1)) 
\end{eqnarray} 
These results enable us to write down the important order parameters
at time $2$: 
\begin{eqnarray} 
m(2)&=&\frac{1}{a}\left\langle\!\left\langle 
     \xi\int Dz\int Dy\, V_{\beta}\big(h'_{1}(z),\theta'_{1}(y)\big)
	       \right\rangle\!\right\rangle 
   \\ 
q(2)&=&\left\langle\!\left\langle 
      \int Dz\int Dy\, W_{\beta}\big(h'_{1}(z),\theta'_{1}(y)\big) 
	      \right\rangle\!\right\rangle 
   \\ 
l(2)&=&\left\langle\!\left\langle 
     \eta\int Dz\int Dy\, W_{\beta}\big(h'_{1}(z),\theta'_{1}(y)\big)
	       \right\rangle\!\right\rangle 
\end{eqnarray} 
defining 
\begin{eqnarray} 
h'_{1}(z)&=&\frac{1}{a}\xi
m(1)+\frac{\alpha}{a}\chi_{h}(0)\sigma(0)+\sqrt{\alpha a D(1)}\,z 
     \\ 
\theta'_{1}(y)&=&\eta
l(1)+\frac{\alpha}{\tilde{a}}\chi_{\theta}(0)
\sigma^{2}(0)+\sqrt{\frac{\alpha}{\tilde
a}E(1)}\,y 
\end{eqnarray} 
The calculation of the variance of the residual overlaps proceeds
as follows. From the recursion relations (\ref{drec})-(\ref{erec})
we get 
\begin{eqnarray} 
D(2)&=&\frac{q(2)}{a^{3}}+\chi_{h}^{2}(1)D(1)+2\chi_{h}(1)(R(2,1) 
	      +\chi_{h}(0)R(2,0)) 
\\ 
E(2)&=&\frac{q(2)}{\tilde{a}}+\chi_{\theta}^{2}(1)E(1) 
		 +2\chi_{\theta}(1)(S(2,1)+\chi_{\theta}(0)S(2,0)) 
\end{eqnarray} 
where 
\begin{eqnarray} 
R(2,0)&=&\frac{1}{a^{3}}\left\langle\!\left\langle 
    \sigma(0)\int Dz\int Dy\,
    V_{\beta}\big(h'_{1}(z),\theta'_{1}(y)\big) 
			\right\rangle\!\right\rangle 
\\ 
S(2,0)&=&\frac{1}{\tilde{a}}\left\langle\!\left\langle 
     \sigma^{2}(0)\int Dz\int Dy\, 
     W_{\beta}\big(h'_{1}(z), \theta'_{1}(y)\big) 
     \right\rangle\!\right\rangle 
\end{eqnarray} 

To obtain the further correlations introduced we remark that because
$\sigma(1)$ appears in the expressions, we have to introduce the
correlation between time steps $0$ and $1$. The relevant correlation
coefficients are given by 
\begin{eqnarray} 
\rho_{h}(t,t')&=& 
\frac{E[(h(t)-M(t))(h(t')-M(t'))]} 
	       {\sqrt{\alpha a D(t)}\sqrt{\alpha a D(t')}} 
\\ 
\rho_{\theta}(t,t')&=& 
\frac{E[(\theta(t)-L(t))(\theta(t')-L(t'))] 
		   } 
	 {\sqrt{({\alpha}/{\tilde a})E(t)} 
		      \sqrt{({\alpha}/{\tilde a})E(t')}} 
\end{eqnarray} 
and the joint distribution 
\begin{equation} 
D\omega_{x}^{a,b}(z,y)=\frac{dzdy}{2\pi\sqrt{1-\rho_{x}^{2}(a,b)}} 
   \exp \left(-\frac{z^{2}-2zy\rho_{x}(a,b)+y^{2}} 
			   {2(1-\rho_{x}^{2}(a,b))}\right) 
\end{equation} 
Then we arrive at 
\begin{align} 
R(2,1)&=\frac{1}{a^{3}}\left\langle\!\left\langle 
   \int D\omega_{h}^{1,0}(z,s)D\omega_{\theta}^{1,0}(y,t)\, 
V_{\beta}\big(h'_{1}(z),\theta'_{1}(y)\big)\, 
    V_{\beta}\big(h'_{0}(s), \theta'_{0}(t)\big) 
  \right\rangle\!\right\rangle 
\\ 
S(2,1)&=\frac{1}{\tilde{a}}\left\langle\!\left\langle 
    \int D\omega_{h}^{1,0}(z,s)D\omega_{\theta}^{1,0}(y,t)\,
    W_{\beta}\big(h'_{1}(z), \theta'_{1}(y)\big)\, 
    W_{\beta}\big(h'_{0}(s), \theta'_{0}(t)\big) 
       \right\rangle\!\right\rangle 
\end{align} 
Finally, the susceptibilities at time step $1$ are given by 
\begin{eqnarray} 
\chi_{h}(1)&=&\frac{1}{a}\left\langle\!\left\langle 
  \int Dz\int Dy \, 
  \left.\Big(\frac{ 
     {\partial V_{\beta}(h(t),\theta(t))}} 
     {{\partial h(t)}}\Big)\right|_{(h'_1(z),\theta'_1(y))} 
    \right\rangle\!\right\rangle 
\\ 
\chi_{\theta}(1)&=&\frac{1}{\tilde{a}}\left\langle\!\left\langle 
    \int Dz\int Dy \, 
   \left.\Big(\frac{ 
     {\partial W_{\beta}(h(t),\theta(t))}} 
     {{\partial \theta(t)}}\Big)\right|_{(h'_1(z),\theta'_1(y))} 
    \right\rangle\!\right\rangle 
\end{eqnarray}

\subsection{Third time step} 

We have all the quantities needed to write down the local fields
at time $t=2$ 
\begin{eqnarray} 
h_{i}(2)&=&\frac{\xi_{i}}{a}m(2) 
 +\frac{\alpha}{a}\chi_{h}(1)(\sigma_{i}(1) 
	 +\chi_{h}(0)\sigma(0))+\mathcal{N}(0,\alpha a D(2)) 
\\ 
\theta_{i}(2)&=&\eta_{i}l(2) 
     +\frac{\alpha}{\tilde{a}}\chi_{\theta}(1)(\sigma_{i}^{2}(1) 
		     +\chi_{\theta}(0)\sigma^{2}(0)) 
	 +\mathcal{N}(0,\frac{\alpha}{\tilde{a}}E(2)) 
\end{eqnarray} 
This gives for the order parameters 
\begin{align} 
m(3)&=\frac{1}{a}\left\langle\!\left\langle 
   \xi\int D\omega_{h}^{2,0}(z,s)D\omega_{\theta}^{2,0}(y,t)\, 
\sum_{\sigma(1)}\,V_{\beta}\big(h'_{2}(z),\theta'_{2}(y)\big)\, 
    U_{\beta}(h'_0(s),\theta'_0(t),\sigma(1)) 
    \right\rangle\!\right\rangle 
\\ 
q(3)&=\left\langle\!\left\langle 
   \int D\omega_{h}^{2,0}(z,s)D\omega_{\theta}^{2,0}(y,t)\,
   \sum_{\sigma(1)}\,W_{\beta}\big(h'_{2}(z),\theta'_{2}(y)\big)\, 
   U_{\beta}(h'_0(s), \theta'_0(t),\sigma(1)) 
   \right\rangle\!\right\rangle 
\\ 
l(3)&=\left\langle\!\left\langle\eta 
  \int D\omega_{h}^{2,0}(z,s)D\omega_{\theta}^{2,0}(y,t)\,
  \sum_{\sigma(1)}\,W_{\beta}\big(h'_{2}(z),\theta'_{2}(y)\big)\, 
  U_{\beta}(h'_0(s), \theta'_0(t),\sigma(1)) 
       \right\rangle\!\right\rangle 
\end{align} 
with 
\begin{equation} 
U_{\beta}(h'_0(s), \theta_0'(t),\sigma(1)) 
=\frac{\exp[\beta\sigma(1)h_{0}'(s)+\beta\sigma^{2}(1)\theta_{0}'(t)]}
{2\exp(\beta\theta_{0}'(t))\cosh(\beta h_{0}'(s))+1} 
\end{equation} 
and 
\begin{eqnarray} 
h'_{2}(z)&=&\frac{\xi}{a}m(2) 
 +\frac{\alpha}{a}\chi_{h}(1)(\sigma(1) 
	 +\chi_{h}(0)\sigma(0))+\sqrt{\alpha a D(2)}\, z 
\\ 
\theta'_{2}(y)&=&\eta l(2) 
     +\frac{\alpha}{\tilde{a}}\chi_{\theta}(1)(\sigma^{2}(1) 
		     +\chi_{\theta}(0)\sigma^{2}(0)) 
	 +\sqrt{\frac{\alpha}{\tilde{a}}E(2)} \, y 
\end{eqnarray} 
In an analogous way further time steps could be calculated.

\section{The complete GFA calculation}

Starting from the path (\ref{eq:transpath}) with the transition
probabilities given by (\ref{eq:transzero}) the generating function
(\ref{genfun}) can be written as 
\begin{align} 
Z[\Psi, \Phi]&= 
\sum_{\bsigma(0)}...\sum_{\bsigma(t)}\mbox{Prob}(\bsigma(0)) 
      \prod_{i=1}^N\prod_{s=1}^{t} 
\exp{\left[-i\sigma_{i}(s)\left(\psi_{i}(s)+ 
	     i\beta h_{i}(\bsigma(s-1))\right)\right]} 
	     \nonumber\\ 
&\quad \times \exp{\left[-i\sigma_{i}^{2}(s)\left(\phi_{i}(s)+ 
	     i\beta \theta_{i}(\bsigma(s-1))\right)\right]} 
   \nonumber\\ 
&\quad \times \exp{\left[-\ln\left(2\exp{(\beta
\theta_{i}(\bsigma(s-1)))}\cosh(\beta
h_{i}(\bsigma(s-1)))+1\right)\right]} 
\end{align} 
where we have left out the subscript $N$. 
We then isolate the local fields by inserting the appropriate delta
distributions 
\begin{align} 
\label{one} 
Z[ \Psi, \Phi]
 &= 
   \sum_{\bsigma(0)}...\sum_{\bsigma(t)}\mbox{Prob}(\bsigma(0)) 
     \int\frac{d{\bf h}d{\bf \hat{h}}d{\bf \theta}d{\bf
       {\hat{\theta}}}}{(2\pi)^{2N(t+1)}} 
   \nonumber\\ 
 &\times
   \prod_{s=1}^{t}\left\{\exp\left[i\sum_{i=1}^{N}\hat{h}_{i}(s)(h_{i}(s)-
     \gamma_{h,i}(s))+i\sum_{i=1}^{N}\hat{\theta}_{i}(s)
       (\theta_{i}(s)-\gamma_{\theta,i}(s))\right]\right.
   \nonumber\\ 
 &\times
   \exp\left[-i\sum_{i=1}^{N}\sigma_{i}(s)(\psi_{i}(s)+i\beta
     h_{i}(s-1))-i\sum_{i=1}^{N}\sigma_{i}^{2}(s)(\phi_{i}(s)+i\beta
     \theta_{i}(s-1))\right] 
   \nonumber\\ 
 &\times
   \left. \exp\left[-\ln\{2\exp{\left(\beta
     \theta_{i}(s-1)\right)}\cosh(\beta h_{i}(s-1))+1\}\right]
     \rule{0cm}{0.6cm}\right\} 
   \nonumber\\ 
 &\times
   \exp{\left[i\sum_{i=1}^{N}\hat{h}_{i}(0)h_{i}(0)
      +i\sum_{i=1}^{N}\hat{\theta}_{i}(0)\theta_{i}(0)\right]}
   \nonumber\\ 
 &\times
   \exp{[\mathcal{F}]} 
\end{align} 
where $\exp{[\mathcal{F}]}$ includes all $J$ and $K$ dependent
terms 
\begin{eqnarray} 
\exp{[\mathcal{F}]}
 &\equiv&
    \exp{\left[-iN\sum_{s=0}^{t}\left(\frac{1}{N}
      \sum_{i=1}^{N}\hat{h}_{i}(s)\right)
        \left(\sum_{j=1}^{N}J_{ij}\sigma_{j}(s)\right)\right]}
    \nonumber\\ 
 &\times&
    \exp{\left[-iN\sum_{s=0}^{t}\left(\frac{1}{N}
      \sum_{i=1}^{N}\hat{\theta}_{i}(s)\right)
         \left(\sum_{j=1}^{N}K_{ij}\sigma_{j}^{2}(s)\right)\right]}
    \nonumber\\ 
 &\times& 
    \exp{\left[i\sum_{s=0}^{t}\sum_{i=1}^{N}
      \left(\hat{h}_{i}(s)J_{ii}\sigma_{i}(s)
        +\hat{\theta}_{i}(s)K_{ii}\sigma_{i}^{2}(s)\right)\right]}
\label{befav} 
\end{eqnarray} 
We remark that in the expression above, the diagonal terms are
excluded by the introduction of the last factor, which we denote
by $M$. 

We first concentrate on the terms containing the disorder, i.e.,
the terms containing $J_{ij}$ and $K_{ij}$, of which we need to
know the limit $N \to \infty$. Using the learning rule (\ref{hebb})
and taking for simplicity one condensed pattern again, say $\mu=1$,
we can study the condensed and the non-condensed part of the noise
(\ref{befav}) by splitting 
\begin{equation} 
\exp{[\mathcal{F}]}=\exp{[\mathcal{F}_c+\mathcal{F}_{nc}]}\,. 
\end{equation} 

With respect to the condensed pattern it is seen that the following
single time variables are relevant 
\begin{eqnarray} 
\label{nene} 
m(s)&\equiv&\frac{1}{aN}\sum_{i=1}^{N}\xi_{i}^{1}\sigma_{i}(s)\qquad
k_h(s)\equiv\frac{1}{aN}\sum_{i=1}^{N}\xi_{i}^{1}\hat{h}_{i}(s)\\ 
\label{nena} 
l(s)&\equiv&\frac{1}{N}\sum_{i=1}^{N}\eta_{i}^{1}\sigma_{i}^{2}(s)\qquad
k_{\theta}(s)\equiv\frac{1}{N}\sum_{i=1}^{N} 
			    \eta_{i}^{1}\hat{\theta}_{i}(s) 
\end{eqnarray} 
To achieve site factorisation we then introduce appropriate delta
distributions for (\ref{nene})-(\ref{nena}) and their conjugated
variables denoted with a hat. Consequently, 
\begin{align} 
&\exp{[\mathcal{F}_c]} 
=\int d{\bf m}d{\bf {\hat{m}}}d{\bf k}_h 
d{\bf {\hat{k}}}_hd{\bf l}d{\bf {\hat{l}}} 
d{\bf k}_{\theta}d{\bf {\hat{k}}}_{\theta} 
\frac{N^{4(t+1)}}{(2\pi)^{4(t+1)}} 
	 \nonumber\\ 
&\times\exp\left\{ 
	       -i N \left( \sprod{k_h}{m} + \sprod{k_{\theta}}{l} ) 
	    \right)\right\} 
    \nonumber\\ 
&\times \exp\left\{-i N \left( 
			 \sprod{\hat{m}}{m} 
			 + \sprod{\hat{k}_h}{k_h} 
			 + \sprod{\hat{l}}{l} 
			 + \sprod{\hat{k}_{\theta}}{k_{\theta}} 
	     \right)\right\} 
       \nonumber\\ 
&\times \exp\left\{-i\sum_{s=0}^{t}\sum_{i=1}^{N} 
    \left(\hat{m}(s)\frac{\xi_{i}}{a}^{1}\sigma_{i}(s) 
       +\hat{k}_h(s)\frac{\xi_{i}}{a}^{1}\hat{h}_{i}(s) 
       +\hat{l}(s)\eta_{i}^{1}\sigma_{i}^{2}(s) 
       +\hat{k}_{\theta}(s)\eta_{i}^{1}\hat{\theta}_{i}(s) 
    \right)\right\} 
\end{align} 

With respect to the non-condensed patterns, we first 
introduce gaussian integration \linebreak[4]
$D{\bf x}\equiv \prod_{s=0}^{t}\left((2\pi)^{-1/2}dx(s) 
\exp{[-\frac{1}{2}x^{2}(s)]}\right)$ to arrive at (we forget about
superindex 1 for the condensed pattern) 
\begin{align} 
\exp{[\mathcal{F}_{nc}]}&=\left\langle\int D{\bf x}_{1}D{\bf
x}_{2}D{\bf y}_{1}D{\bf y}_{2}\right.\nonumber\\ 
&\times\exp\left\{\frac{1}{\sqrt{2iN}}\sum_{i=1}^{N} 
\frac{\xi_{i}}{a}\sum_{s=0}^{t}\left(x_{1}(\hat{h}_{i}(s)+\sigma_{i}(s))
	+iy_{1}(s)(\hat{h}_{i}(s)-\sigma_{i}(s))\right)\right\} 
	\nonumber\\ 
&\times\left. \exp\left\{\frac{1}{\sqrt{2iN}}\sum_{i=1}^{N} 
\eta_{i}\sum_{s=0}^{t}\left(x_{2}(\hat{\theta}_{i}(s)+\sigma_{i}^{2}(s))
+iy_{2}(s)(\hat{\theta}_{i}(s)-\sigma_{i}^{2}(s)) 
     \right)\right\}\right\rangle_{\xi_i}^{p-1} 
     \nonumber\\ 
&\times \exp\left\{i\alpha\sum_{s,s'=0}^{t}\sum_{i=1}^{N} 
     \left(\frac{1}{a}\hat{h}_{i}(s')\sigma_{i}(s) 
	+\frac{1}{\tilde{a}}\hat{\theta}_{i}(s')\sigma_{i}^{2}(s) 
	   \right)\delta_{ss'}\right\} 
\end{align} 
with clear notation. We have also rewritten $M$ in a more suitable
way. Performing the average over the patterns and expanding the
exponentials up to order $O(N^{-1})$ it is seen that the following
two time variables turn up 
\begin{eqnarray} 
q(s,s')&\equiv& 
    \frac{1}{Na}\sum_{i=1}^{N}\sigma_{i}(s)\sigma_{i}(s') 
\qquad Q_h(s,s')\equiv 
     \frac{1}{Na}\sum_{i=1}^{N}\hat{h}_{i}(s)\hat{h}_{i}(s') 
       \\ 
p(s,s')&\equiv& 
   \frac{1}{N\tilde{a}}\sum_{i=1}^{N}\sigma_{i}^{2}(s)\sigma_{i}^{2}(s')
\qquad Q_{\theta}(s,s')\equiv \frac{1}{N\tilde{a}} 
      \sum_{i=1}^{N}\hat{\theta}_{i}(s)\hat{\theta}_{i}(s') 
	\\ 
K_h(s,s')&\equiv& 
    \frac{1}{Na}\sum_{i=1}^{N}\sigma_{i}(s)\hat{h}_{i}(s') 
    \qquad K_{\theta}(s,s')\equiv 
\frac{1}{N\tilde{a}}\sum_{i=1}^{N}\sigma_{i}^{2}(s)\hat{\theta}_{i}(s')
\end{eqnarray} 
Introducing again appropriate delta distributions we obtain after
some matrix algebra 
\begin{align} 
&\exp{[\mathcal{F}_{nc}]}=\left(\frac{N}{2\pi}\right)^{12(t+1)}\int
d{\bf q}d{\bf{\hat{q}}}d{\bf Q}_h 
d{\bf {\hat{Q}}}_hd{\bf K}_hd{\bf {\hat{K}}}_h 
d{\bf p}d{\bf {\hat{p}}}d{\bf Q}_{\theta} 
d{\bf {\hat{Q}}}_{\theta}d{\bf K}_{\theta} 
d{\bf {\hat{K}}}_{\theta} 
      \nonumber\\ 
&\times \exp 
      \left[iN \mbox{Tr} \left( 
	  \smprod{\hat{q}}{q} 
	+ \smprod{\hat{Q}_h}{Q_h}+ \smprod{\hat{K}_h}{K_h} 
	+ \smprod{\hat{p}}{p} + \smprod{\hat{Q}_{\theta}}{Q_{\theta}}
	+ \smprod{\hat{K}_{\theta}}{K_{\theta}} 
      \right)\right] 
\nonumber\\ 
&\times \exp \left[i\alpha N 
		\mbox{Tr} 
		\left( 
		   {\bf K_h} + {\bf K_{\theta}}\right) 
		\right] 
    \nonumber\\ 
&\times \exp \left[\frac{-i}{a}\sum_{s,s'}^{t}\sum_{i=1}^{N} 
\left(\hat{q}(s,s')\sigma_{i}(s)\sigma_{i}(s') 
   +\hat{Q}_h(s,s')\hat{h}_{i}(s)\hat{h}_{i}(s') 
   +\hat{K}_h(s,s')\sigma_{i}(s)\hat{h}_{i}(s')\right)\right] 
\nonumber\\ 
&\times \exp \left[\frac{-i}{\tilde{a}}\sum_{s,s'}^{t}\sum_{i=1}^{N}
\left(\hat{p}(s,s')\sigma_{i}^{2}(s)\sigma_{i}^{2}(s') 
     +\hat{Q}_{\theta}(s,s')\hat{\theta}_{i}(s)\hat{\theta}_{i}(s') 
     +\hat{K}_{\theta}(s,s')\sigma_{i}^{2}(s)\hat{\theta}_{i}(s') 
	  \right)\right] 
\nonumber\\ 
&\times \exp {\left[-\frac{p}{2}\left[\ln\left({\bf Q}_h{\bf
q}-(i{\bf I}-{\bf K}_h^{\dagger})(i{\bf I}-{\bf K}_h)\right) 
	+\ln\left({\bf Q}_{\theta}{\bf p}-(i{\bf I}-{\bf
	K}_{\theta}^{\dagger})(i{\bf I}-{\bf K}_{\theta}) 
		\right)\right]\right]}\nonumber\\ 
\end{align} 
with ${\bf I}$ the unit matrix of dimension $(t+1)$.

This concludes the discussion of the averaging over the disorder with
as result a generating functional of the form $Z\sim\exp{N...}$ which
can be evaluated in the limit $N \to \infty$ via the saddle-point
method. 
\begin{equation} 
\label{two} 
\overline {Z[\Psi,\Phi]}\propto 
\int \{ \dots \} 
\exp \left[ N \left(
 \alpha \ln \Omega 
 + \Xi
 + \Lambda \right)
 \right]
\end{equation}
%&&
%\int d{\bf m}d{\bf \hat{m}}d{\bf k}_hd{\bf \hat{k}}_h 
%d{\bf l}d{\bf \hat{l}}d{\bf k}_{\theta} 
%d{\bf \hat{k}}_{\theta} 
%d{\bf q}d{\bf {\hat{q}}} 
%d{\bf Q}_hd{\bf {\hat{Q}}}_h 
% \nonumber\\ 
%d{\bf K}_hd{\bf {\hat{K}}}_h 
%d{\bf p}d{\bf {\hat{p}}}d{\bf Q}_{\theta} 
%d{\bf {\hat{Q}}}_{\theta}d{\bf K}_{\theta} 
%d{\bf {\hat{K}}}_{\theta} 
%	\nonumber\\
%\\
%&&\times 
%\exp\left\{\alpha N \ln\left[\Omega({\bf q},{\bf Q_h},{\bf K_h},{\bf
%p}, 
%  {\bf Q_\theta},{\bf K_\theta})\right]\right\} 
%	   \nonumber\\ 
%&&\times 
%\exp\left\{ N\Xi({\bf m},{\bf \hat{m}},{\bf k}_h,{\bf \hat{k}}_h, 
%{\bf l},{\bf \hat{l}},{\bf k}_{\theta},{\bf \hat{k}}_{\theta}, 
%{\bf q},{\bf {\hat{q}}},{\bf Q}_h,{\bf \hat{Q}}_h,{\bf K}_h, 
%{\bf {\hat{K}}}_h,{\bf p},{\bf {\hat{p}}},{\bf Q}_{\theta}, 
%{\bf{\hat{Q}}}_{\theta},{\bf K}_{\theta},{\bf {\hat{K}}}_{\theta}) 
%\right\} 
%	 \nonumber\\ 
%&&\times 
%\exp\left\{ N\Lambda ({\bf m},{\bf \hat{m}},{\bf k}_h,{\bf
%\hat{k}}_h, 
%{\bf l},{\bf \hat{l}},{\bf k}_{\theta},{\bf \hat{k}}_{\theta}, 
%{\bf {\hat{q}}},{\bf {\hat{Q}}}_h, 
%{\bf {\hat{K}}}_h,{\bf {\hat{p}}}, 
%{\bf{\hat{Q}}}_{\theta},{\bf {\hat{K}}}_{\theta})+O(\ln N) 
%\right\} 
%\end{eqnarray} 
with 
\begin{equation} 
\Omega=\left([{\bf Q}_h {\bf q}-(i{\bf I}-{\bf K}_h^{\dagger})(i{\bf
I}-{\bf K}_h)][{\bf Q}_\theta {\bf p}-(i{\bf I}-{\bf
K}_\theta^{\dagger})(i{\bf I}-{\bf K}_\theta)]\right)^{-\frac{1}{2}}
\end{equation} 
\begin{multline} 
\Xi 
  =  \ 
  i \left( 
	    \sprod{\hat{m}}{m} 
	  + \sprod{\hat{k}_h}{k_h} + \sprod{k_h}{m} 
	  + \sprod{\hat{l}}{l} 
	  + \sprod{\hat{k}_\theta}{k_\theta} + \sprod{k_\theta}{l} 
	 \right) \\ 
  + 
  i \mbox{Tr} 
	\left( 
	    \smprod{\hat{q}}{q} 
	  + \smprod{\hat{Q}_h}{Q_h} 
	  + \smprod{\hat{K}_h}{K_h} 
	  + \smprod{\hat{p}}{p} 
	  + \smprod{\hat{Q}_\theta}{Q_\theta} 
	  + \smprod{\hat{K}_\theta}{K_\theta} 
	  + \alpha ({\bf K_h} + {\bf K_\theta}) 
    \right) 
\end{multline} 
\begin{align} 
&\Lambda
    =\frac{1}{N}\sum_{i=1}^{N}\ln\left\{\int
      d{\bf h} d{\bf \hat{h}} d{\boldsymbol \theta}
      d{\hat{\boldsymbol\theta}}\sum_{\bsigma(0)...\bsigma(t)}
      \mbox{Prob}(\bsigma(0))\right.
    \nonumber\\
&\times
     \exp{\left[i\sum_{s=0}^{t}\hat{h}_{i}(s)(h_{i}(s)
       -\gamma_{h,i}(s)-\hat{k}_h(s)\frac{\xi_{i}}{a})\right]}
     \nonumber\\
&\times
     \exp{\left[i\sum_{s=0}^{t}\hat{\theta}_{i}(s)(\theta_{i}(s)
       -\gamma_{\theta,i}(s)-\hat{k}_\theta (s)\eta_{i})
       -\sum_{s=1}^{t}\ln{\left(2\exp{\left(\beta\theta(s-1)\right)}
         \cosh(\beta h(s-1))+1\right)}\right]}
  \nonumber\\ 
&\times
     \exp{\left[-i\sum_{s=0}^{t}
       \left(\sigma(s)(\psi_{i}(s)+\hat{m}(s)\frac{\xi_{i}}{a}
               +i\beta h(s-1))+\sigma^{2}(s)(\phi_{i}(s)+\hat{l}(s)\eta_{i}
	       +i\beta \theta(s-1))\right)\right]}
  \nonumber\\ 
&\times
     \exp{\left[-\frac{i}{a}\sum_{s,s'=0}^{t}
       \hat{q}(s,s')\sigma(s)\sigma(s')
       +\hat{K}_h(s,s')\sigma(s)\hat{h}(s')
       +\hat{Q}_h(s,s')\hat{h}(s)\hat{h}(s')\right]}
  \nonumber\\
&\times
     \left.
       \exp{\left[-\frac{i}{\tilde{a}}
         \sum_{s,s'=0}^{t}
	   \hat{p}(s,s')\sigma^{2}(s)\sigma^{2}(s')
	   +\hat{K}_\theta(s,s')\sigma^{2}(s)\hat{\theta}(s')
	   +\hat{Q}_\theta(s,s')\hat{\theta}(s)\hat{\theta}(s')
            \right]}
     \right\}
\end{align} 

At this point, some remarks are in order. First, we have been
able to factorize the generating functional with respect to
the sites. Further, the external fields are only present in
$\Lambda$. Therefore it is useful to introduce the effective single
spin measure 
\begin{equation} 
\langle f\rangle_{*,i}
  =\frac{\int d{\bf h}d{\bf \hat{h}}
              d{\boldsymbol\theta}
	      d{\hat{\boldsymbol\theta}} 
           \sum_{\sigma(0)...\sigma(t)}
	         \mbox{Prob}({\sigma}(0))
		 \mathcal{M}_{i}
		   \left[
		     \bsigma,{\bf \xi}_{i},{\bf h},
		     {\bf \hat{h}},{\boldsymbol \theta},
		     {\hat{\boldsymbol\theta}}
		   \right]
		 f}
        {\int d{\bf h}d{\bf \hat{h}}
	      d{\boldsymbol \theta}
	      d{\hat{\boldsymbol \theta}}
	   \sum_{\sigma(0)...\sigma(t)}
	         \mbox{Prob}({\sigma}(0))
		 \mathcal{M}_{i}
		   \left[
		     \bsigma,{\bf \xi}_{i},{\bf h},
		     {\bf \hat{h}},{\boldsymbol \theta},
		     {\hat{\boldsymbol\theta}}
		   \right]}
\label{spinmeas} 
\end{equation} 
where the auxiliary fields $\psi_{i}$ and $\phi_{i}$ have been
taken to be zero and $\mathcal{M}_{i}$ is given by 
\begin{align} 
\mathcal{M}_{i}[...]
  &= \exp{\left[i\sum_{s=0}^{t}\hat{h}(s)(h(s)
        -\gamma_h(s)-m(s)\frac{\xi_{i}}{a})
	+i\sum_{i=1}^{N}\hat{\theta}(s)(\theta(s)
	-\gamma_\theta(s)-l(s)\eta_{i})\right]}
      \nonumber\\
  &\times
     \exp{\left[-\sum_{s=1}^{t}
       \ln{\left(2\exp{\left(\beta\theta(s-1)\right)}
           \cosh(\beta h(s-1))+1\right)}\right]}
     \nonumber \\ 
  &\times
     \exp{\left[-i\sum_{s=1}^{t}
       \sigma(s)\left(i\beta h(s-1)+k_h(s)\frac{\xi_{i}}{a}\right)
       +\sigma^{2}(s)\left(i\beta\theta(s-1)+k_\theta(s)\eta_{i}
       \right)\right]}
   \nonumber\\ 
  &\times
     \exp{\left[-i\sigma(0)k_h(0)\frac{\xi_{i}}{a}
          -i\sigma^{2}(0)k_\theta (0)\eta_{i}\right]}
     \nonumber\\ 
  &\times
     \exp{\left[-\frac{i}{a}
        \sum_{s,s'}^{t}\hat{q}(s,s')\sigma(s)\sigma(s')
	+\hat{Q}_h(s,s')\hat{h}(s)\hat{h}(s')
	+\hat{K}_h(s,s')\sigma(s)\hat{h}(s')
	\right]}
     \nonumber\\
  &\times
     \exp{\left[-\frac{i}{\tilde{a}}
        \sum_{s,s'}^{t}\left(\hat{p}(s,s')\sigma^{2}(s)\sigma^{2}(s')
	+\hat{Q}_\theta(s,s')\hat{\theta}(s)\hat{\theta}(s')
	+\hat{K}_\theta(s,s')\sigma^{2}(s)\hat{\theta}(s')
	\right)\right]}
\end{align} 
The parameters ${{\bf m}, {\bf k}_h, {\bf l}, {\bf k}_\theta,
\hat{\bf q},\hat{\bf Q}_h,\hat{\bf K}_h,\hat{\bf p},
\hat{\bf Q}_\theta,\hat{\bf K}_\theta }$ in $\mathcal{M}_{i}[...]$ 
are defined as the solutions of the saddle-point equation
$d(\Xi+\Lambda)=0$. Furthermore, we have dropped the subindices 
$i$ from the external fields since they are taken to be
site-independent. 

Before determining the saddle-point solutions we write down the
following identities 
\begin{equation} 
\lim_{\Psi,\Phi\rightarrow 0}\frac{\partial
\overline{Z[\Psi,\Phi]}}{\partial 
\psi_{i}(s)}=\lim_{\Psi,\Phi\rightarrow 0}\frac{\int d{\bf 
m}...d{\bf\hat{K}_\theta}e^{N(\Xi+\Lambda)+\alpha
N\ln{(\Omega)}+O(\ln{(N)})}\left[N\frac{\partial 
\Lambda}{\partial \psi_{i}(s)}\right]}{\int d{\bf 
m}...d{\bf\hat{K}_\theta}e^{N(\Xi+\Lambda)+\alpha 
N\ln{(\Omega)}+O(\ln{(N)})}}=-i\langle\sigma(s)\rangle_{*,i} 
\end{equation} 
\begin{equation} 
\lim_{\Psi,\Phi\rightarrow 0}\frac{\partial
\overline{Z[\Psi,\Phi]}}{\partial 
\phi_{i}(s)}=-i\langle\sigma^{2}(s)\rangle_{*,i} 
\end{equation} 
\begin{equation} 
\lim_{\Psi,\Phi\rightarrow 0}\frac{\partial^{2}
\overline{Z[\Psi,\Phi]}}{\partial 
\psi_{i}(s)\partial\psi_{j}(s')}=-\delta_{ij}
[\langle\sigma(s)\sigma(s')\rangle_{*,i}
-\langle\sigma(s)\rangle_{*,i}\langle\sigma(s')
\rangle_{*,i}]-\langle\sigma(s)\rangle_{*,i}\langle\sigma(s')\rangle_{*,j}
\end{equation} 
\begin{equation} 
\lim_{\Psi,\Phi\rightarrow 0}\frac{\partial^{2}
\overline{Z[\Psi,\Phi]}}{\partial 
\phi_{i}(s)\partial\phi_{j}(s')}
=-\delta_{ij}[\langle\sigma^{2}(s)\sigma^{2}(s')\rangle_{*,i}
-\langle\sigma^{2}(s)\rangle_{*,i}\langle\sigma^{2}(s')\rangle_{*,i}]
-\langle\sigma^{2}(s)\rangle_{*,i}\langle\sigma^{2}(s')\rangle_{*,j}
\end{equation} 
\begin{equation} 
\lim_{\Psi,\Phi\rightarrow 0}\frac{\partial^{2}
\overline{Z[\Psi,\Phi]}}{\partial 
\psi_{i}(s)\partial\phi_{j}(s')}
=-\delta_{ij}[\langle\sigma(s)\sigma^{2}(s')\rangle_{*,i}
-\langle\sigma(s)\rangle_{*,i}\langle\sigma^{2}(s')\rangle_{*,i}]
-\langle\sigma(s)\rangle_{*,i}\langle\sigma^{2}(s')\rangle_{*,j}
\end{equation} 
\begin{equation} 
\lim_{\Psi,\Phi\rightarrow 0}\frac{\partial^{2}
\overline{Z[\Psi,\Phi]}}{\partial 
\psi_{i}(s)\partial\gamma_{h,j}(s')}
=-\delta_{ij}[\langle\sigma(s)\hat{h}(s')\rangle_{*,i}
-\langle\sigma(s)\rangle_{*,i}\langle\hat{h}(s')\rangle_{*,i}]
-\langle\sigma(s)\rangle_{*,i}\langle\hat{h}(s')\rangle_{*,j}
\end{equation} 
\begin{equation} 
\lim_{\Psi,\Phi\rightarrow 0}\frac{\partial^{2}
\overline{Z[\Psi,\Phi]}}{\partial 
\phi_{i}(s)\partial\gamma_{\theta,j}(s')}
=-\delta_{ij}[\langle\sigma^{2}(s)\hat{\theta}(s')\rangle_{*,i}
-\langle\sigma^{2}(s)\rangle_{*,i}\langle\hat{\theta}(s')\rangle_{*,i}]
-\langle\sigma^{2}(s)\rangle_{*,i}\langle\hat{\theta}(s')\rangle_{*,j}
\end{equation} 
\begin{equation} 
\lim_{\Psi,\Phi\rightarrow 0}\frac{\partial\overline{Z[\Psi,\Phi]}}
{\partial\gamma_{h,i}(s)}=-i\langle\hat{h}(s)\rangle_{*,i}
\end{equation} 
\begin{equation} 
\label{una} \lim_{\Psi,\Phi\rightarrow 0}\frac{\partial
\overline{Z[\Psi,\Phi]}}{\partial\gamma_{\theta,i}(s)}
=-i\langle\hat{\theta}(s)\rangle_{*,i}
\end{equation} 
where we recall that the overline denotes the disorder average
and where we used the normalization $\overline{Z[0,0]}=1$, we
find that the last two equations are equal to $0$, since the limit
$\Psi,\Phi\rightarrow 0$ can be interchanged with the derivative,
due to the fact that the local fields are completely unrelated with
the external fields. 
This leads to $\langle\hat{h}(s)\rangle_{*,i}=0$, 
$\langle\hat{\theta}(s)\rangle_{*,i}=0$,
$\langle\hat{h}(s)\hat{h}(s')\rangle_{*,i}=0$ and 
$\langle\hat{\theta}(s)\hat{\theta}(s')\rangle_{*,i}=0$. 

Recalling (\ref{cordef3}) 
we then have obtained the following identities 
\begin{eqnarray} 
\overline{\langle\sigma_{i}(s)\rangle}
&=&\langle\sigma(s)\rangle_{*,i}\\
\overline{\langle\sigma_{i}^{2}(s)\rangle}
&=&\langle\sigma^{2}(s)\rangle_{*,i}\\
\overline{C_{h,ij}(s,s')}
&=&\delta_{ij}\langle\sigma(s)\sigma(s')\rangle_{*,i}
+(1-\delta_{ij})\langle\sigma(s)\rangle_{*,i}\langle\sigma(s')\rangle_{*,j}\\
\overline{C_{\theta,ij}(s,s')}
&=&\delta_{ij}\langle\sigma^{2}(s)\sigma^{2}(s')\rangle_{*,i}
+(1-\delta_{ij})\langle\sigma^{2}(s)
\rangle_{*,i}\langle\sigma^{2}(s')\rangle_{*,j}\\
\overline{G_{h,ij}(s,s')}
&=&\delta_{ij}\langle\sigma(s)\hat{h}(s')\rangle_{*,i}\\
\overline{G_{\theta,ij}(s,s')}
&=&\delta_{ij}\langle\sigma^{2}(s)\hat{\theta}(s')\rangle_{*,i}
\end{eqnarray} 
from which we obtain by using the LLN 
\footnotesize 
\begin{align} 
C_{h}(s,s')&\equiv \lim_{N\rightarrow\infty}
  \frac{1}{N}\sum_{i=1}^{N}\overline{C_{h,ii}(s,s')}
  =\lim_{N\rightarrow\infty}\frac{1}{N}\sum_{i=1}^{N}
  \langle\sigma(s)\sigma(s')\rangle_{*,i}
  =\left\langle\langle\sigma(s)\sigma(s')
  \rangle_{*}\right\rangle_{\xi}\\
C_{\theta}(s,s')&\equiv \lim_{N\rightarrow\infty}
  \frac{1}{N}\sum_{i=1}^{N}\overline{C_{\theta,ii}(s,s')}
  =\lim_{N\rightarrow\infty}\frac{1}{N}
  \sum_{i=1}^{N}\langle\sigma^{2}(s)\sigma^{2}(s')\rangle_{*,i}
  =\left\langle\langle\sigma^{2}(s)\sigma^{2}(s')\rangle_{*}
  \right\rangle_{\xi}\\
G_{h}(s,s')&\equiv \lim_{N\rightarrow\infty}
  \frac{1}{N}\sum_{i=1}^{N}\overline{G_{h,ii}(s,s')}
  =\lim_{N\rightarrow\infty}\frac{1}{N}\sum_{i=1}^{N}
     \frac{\partial}{\partial\gamma_{h,i}(s')}
     \langle\sigma(s)\rangle_{*,i}
  =-i\left\langle\langle\sigma(s)\hat{h}(s')
     \rangle_{*}\right\rangle_{\xi}\\
G_{\theta}(s,s')&\equiv \lim_{N\rightarrow\infty}
  \frac{1}{N}\sum_{i=1}^{N}\overline{G_{\theta,ii}(s,s')}
  =\lim_{N\rightarrow\infty}\frac{1}{N}
    \sum_{i=1}^{N}\frac{\partial}{\partial\gamma_{\theta,i}(s')}
    \langle\sigma^{2}(s)\rangle_{*,i}
  =-i\left\langle\langle\sigma^{2}(s)\hat{\theta}(s')
     \rangle_{*}\right\rangle_{\xi}
\end{align} 
\normalsize 
where $\langle\cdot\rangle_{\xi}$ denotes the average over
the condensed pattern. Finally, we write down the saddle-point
equations. Variation with respect to ${\bf m}, {\bf k}_h, {\bf l}$
and $ {\bf k}_{\theta}$ gives 
\begin{equation} 
\hat{m}(s)=k_h(s)\qquad\hat{k}_h(s)=m(s)\qquad\hat{l}(s)=k_\theta(s)\qquad\hat{k}_\theta(s)=l(s)
\end{equation} 

Next, we calculate the variation with respect to ${\bf \hat{m}},
{\bf \hat{k}}_h, {\bf \hat{l}}$ and $ {\bf \hat{k}}_{\theta}$,
leading to 
\begin{eqnarray} 
m(s)&=&\left\langle\xi\langle\sigma(s)\rangle_{*}\right\rangle_{\xi}\quad
k_h(s)=\left\langle\xi\langle\hat{h}(s)\rangle_{*}\right\rangle_{\xi}=0
\\ 
l(s)&=&\left\langle\eta\langle\sigma^{2}(s)\rangle_{*}\right\rangle_{\xi}\quad
k_\theta(s)=\left\langle\eta\langle\hat{\theta}(s)\rangle_{*}\right\rangle_{\xi}=0
\end{eqnarray}

Variation with respect to ${\bf \hat{q}}$, ${\bf\hat{Q}}_h$ and
${\bf\hat{K}}_h$ gives 
\begin{equation} 
q(s,s')=\frac{1}{a}C_{h}(s,s')\qquad Q_h(s,s')=0\qquad
K_h(s,s')=\frac{i}{a}G_{h}(s,s') 
\end{equation} 
and with respect to ${\bf\hat{p}}$, ${\bf \hat{Q}}_\theta$ and
${\bf \hat{K}}_\theta$ 
\begin{equation} 
p(s,s')=\frac{1}{\tilde{a}}C_{\theta}(s,s')\qquad
Q_\theta(s,s')=0\qquad
K_\theta(s,s')=\frac{i}{\tilde{a}}G_{\theta}(s,s') 
\end{equation} 

Finally, variation with respect to of ${\bf q}$, ${\bf Q}_h$, ${\bf
K}_h$, ${\bf p}$, ${\bf Q}_\theta$ and ${\bf K}_\theta$ i leads to,
after some algebra 
\begin{eqnarray} 
\hat{q}(s,s')&=&0\qquad\hat{p}(s,s')=0\\ 
\hat{Q}_h(s,s')&=&-\frac{i\alpha a}{2}\left[\left(a{\bf
I}-{\bf G}_{h}\right)^{-1}{\bf C}_{h}\left(a{\bf I}-{\bf
G}_{h}^{\dagger}\right)^{-1}\right](s,s')\\ 
\hat{K}_h(s,s')&=&\alpha a\left[\left(a{\bf I}-{\bf
G}_{h}\right)^{-1}\right](s,s')-\alpha{\bf I}(s,s')\\ 
\hat{Q}_\theta(s,s')&=&-\frac{i\alpha
\tilde{a}}{2}\left[\left(\tilde{a}{\bf I}-{\bf
G}_{\theta}\right)^{-1}{\bf C}_{\theta}\left(\tilde{a}{\bf I}-{\bf
G}_{\theta}^{\dagger}\right)^{-1}\right](s,s')\\ 
\hat{K}_\theta(s,s')&=&\alpha\tilde{a}\left[\left(\tilde{a}{\bf
I}-{\bf G}_{\theta}\right)^{-1}\right](s,s')-\alpha{\bf I}(s,s') 
\end{eqnarray}

After the introduction of the saddle-point solutions, the single
spin measure (\ref{spinmeas}) is expressed in terms of the physical
order parameters, 
$\{{\bf m}$ , ${\bf l}$, ${\bf C}_{h}$, ${\bf C}_{\theta}$,
${\bf G}_{h}$, ${\bf G}_{\theta}\}$. Rewriting the average in the following way 
\begin{multline} 
\langle\!\langle
f[\sigma(0),...,\sigma(t)]\rangle\!\rangle\equiv\left\langle\int
dh(0)...dh(t-1)d\theta(0)...d\theta(t-1)\nonumber\right.\\ 
\left.\sum_{\sigma(0)...\sigma(t)}
\mbox{Prob}\left[h(0),...,h(t-1),\theta(0),...,\theta(t-1);
\sigma(0),...,\sigma(t)\right]
f\left[\sigma(0),...,\sigma(t)\right]\right\rangle_{\xi}
\end{multline} 
with 
\begin{align} 
\label{four} 
&\mbox{Prob}\left[...\right]\sim\mbox{Prob}(\sigma(0))
   \int\prod_{s=0}^{t-1}dz_{h}(s)dz_{\theta}(s)\nonumber\\
&\times\delta\left[h(s)-\gamma_h(s)-m(s)\xi
    -\alpha\sum_{s'=0}^{t}
    \left[\left(a{\bf I}-{\bf G}_{h}\right)^{-1}\right](s,s')\sigma(s')
    +\frac{\alpha}{a}\sigma(s)-\sqrt{\alpha}z_{h}(s)\right]\nonumber\\
&\times\delta\left[\theta(s)-\gamma_\theta(s)-l(s)\eta
    -\alpha\sum_{s'=0}^{t}\left[\left(\tilde{a}{\bf I}
    -{\bf G}_{\theta}\right)^{-1}\right](s,s')\sigma^{2}(s')
    +\frac{\alpha}{\tilde{a}}\sigma^{2}(s)
    -\sqrt{\alpha}z_{\theta}(s)\right]\nonumber\\
&\times\frac{\exp{\left(\beta\sigma(s+1)h(s)
    +\beta\sigma^{2}(s+1)\theta(s)\right)}}
    {2\exp{\left(\beta\theta(s)\right)}\cosh(\beta h(s))+1}\nonumber\\ 
&\times\int d{\bf\hat{h}}d{\hat{\boldsymbol\theta}}
    \exp{\left[-\frac{\alpha}{2}\sum_{s,s'=0}^{t}\left[\left(a{\bf I}
    -{\bf G}_{h}\right)^{-1}{\bf C}_{h}\left(a{\bf I}
    -{\bf G}_{h}^{\dagger}\right)^{-1}\right](s,s')
       \hat{h}(s)\hat{h}(s')\right]}\nonumber\\
&\times\exp{\left[-\frac{\alpha}{2}
    \sum_{s,s'=0}^{t}\left[\left(\tilde{a}{\bf I}
    -{\bf G}_{\theta}\right)^{-1}{\bf C}_{\theta}
    \left(\tilde{a}{\bf I}
          -{\bf G}_{\theta}^{\dagger}\right)^{-1}
    \right](s,s')\hat{\theta}(s)\hat{\theta}(s')\right]}\nonumber\\
&\times\exp{\left[i\sqrt{\alpha}
    \sum_{s=0}^{t}\left(\hat{h}(s)z_{h}(s)
    +\hat{\theta}(s)z_{\theta}(s)\right)\right]}
\end{align} 
the integration over $\hat{\bf h}$ and $\hat{\boldsymbol \theta}$ is easy since
the exponents 
multiplying the terms $\hat{h}(s)\hat{h}(s')$ and
$\hat{\theta}(s)\hat{\theta}(s')$ are symmetric matrices. Doing
the integration and including the normalization factors, we find
the results of Section 4.

\section{First two time steps in the GFA} 

We use the GFA results to calculate the first two time steps of the
dynamics explicitly. The initial conditions are chosen as follows:
$\gamma(0)=\beta(0)=0$, i.e., no external field, 
 $m(0)=m_{0}$, $l(0)=l_{0}$, $C_{h}(0,0)=C_{\theta}(0,0)=q_{0}$, 
$G_{h}(0,0)=G_{\theta}(0,0)=G_{h}(0,1)=G_{\theta}(0,1)=0$.  

\subsection{First time step} 

For time 1, the relevant probabilities are 
\begin{align} 
\label{meas2} 
w[z_{h}(0),z_{\theta}(0)]&=\frac{a\tilde{a}}{2\pi
q_{0}}\exp{\left[-\frac{a^2}{2q_{0}}(z_{h}(0))^{2}
-\frac{\tilde{a}^2}{2q_{0}}(z_{\theta}(0))^{2}\right]}\\
\mbox{Prob}[\sigma(0),
\sigma(1)|z_{h}(0),z_{\theta}(0)]&=\mbox{Prob}(\sigma(0))
\frac{\exp{\left(\beta\sigma(1)h(0)
+\beta\sigma^{2}(1)\theta(0)\right)}}
{2\exp{\left(\beta\theta(0)\right)}\cosh(\beta h(0))+1} 
\end{align} 
with $h(0)={\xi}m_{0}/a + \sqrt{\alpha}z_{h}(0)$ and 
$\theta(0)=l_{0}\eta+\sqrt{\alpha}z_{\theta}(0)$. Then, using
(\ref{cata}) and (\ref{cata2}), 
\begin{align} 
m(1)=\frac{1}{a}\left\langle\!\left\langle\xi\sigma(1)
\right\rangle\!\right\rangle&=
\int\int
Dz_{h}(0)Dz_{\theta}(0)\left\{
\frac{\exp{\left(\beta\theta^{+}(0)\right)}\sinh(\beta
h^{+}(0))}{2\exp{\left(\beta\theta^{+}(0)\right)}\cosh(\beta
h^{+}(0))+1}\right.\nonumber\\ 
&\qquad +\left.\frac{\exp{\left(\beta\theta^{+}(0)\right)}\sinh(\beta
h^{-}(0))}{2\exp{\left(\beta\theta^{+}(0)\right)}\cosh(\beta
h^{-}(0))+1}\right\} 
\end{align} 
with, as usual, $Dz$ a gaussian measure defined in (\ref{meas2})
and where 
$h^{\pm}(0)=\pm {m_{0}}/{a}+\sqrt{\alpha}z_{h}(0)$ and 
$\theta^{+}(0)={l_{0}}/{a}+\sqrt{\alpha}z_{\theta}(0)$. In the
same way we find the values for 
$l(1)$ and $q(1)$, i.e. 
$l(1)=\left\langle\!\left\langle\eta\sigma^{2}(1)\right\rangle\!\right\rangle$
and $q(1)=\left\langle\!\left\langle\sigma^{2}(1)\right\rangle\! 
\right\rangle$. 

The correlation and response functions are 
\begin{eqnarray} 
C_{h}(1,0)&=&C_{h}(0,1)=\left\langle\!\left\langle\sigma(0)
  \sigma(1)\right\rangle\!\right\rangle\\
C_{\theta}(1,0)&=&C_{\theta}(0,1)=\left\langle\!\left\langle\sigma(0)
  \sigma(1)\right\rangle\!\right\rangle\\
G_h(1,0)&=&\beta\left\langle\!\left\langle\sigma(1)\left[\sigma(1)-
V_{\beta}(h(0), \theta(0))\right]\right\rangle\!\right\rangle\\ 
G_{\theta}(t,t')&=&\beta\left\langle\!\left\langle 
\sigma^{2}(1)\left[\sigma^{2}(1)-W_{\beta}(h(0), \theta(0)) 
\right]\right\rangle\!\right\rangle 
\end{eqnarray}

\subsection{Second time step} 

In this time step correlations start to appear. 
\begin{displaymath} 
{\bf C}_{h,\theta}^{t=2}=\left(\begin{array}{cc} q_{0} &
C_{h,\theta}(1,0) \\ C_{h,\theta}(1,0) & q(1) 
\end{array}\right) 
\end{displaymath} 
\begin{displaymath} 
\left(a{\bf I}-{\bf
G}_{h}^{t=2}\right)^{-1}=\frac{1}{a^2}\left(\begin{array}{cc} a &
0 \\ G_{h}(1,0) & a 
\end{array}\right)\qquad\left(\tilde{a}{\bf I}-{\bf
G}_{\theta}^{t=2}\right)^{-1}=\frac{1}{\tilde{a}^2}\left(\begin{array}{cc}
\tilde{a} & 0 \\ 
G_{h}(1,0) & \tilde{a} \end{array}\right) 
\end{displaymath} 
The explicit single-site average for time $2$ is determined by 
\begin{align} 
w[z_{h}(0),z_{h}(1),z_{\theta}(0)z_{\theta}(1)]&
=\frac{1}{4\pi^2\sqrt{(q_0A_h-B_h^2)(q_0A_\theta-B_\theta^2)}}\nonumber\\
&\times\exp{\left[-\frac{1}{2C_h}(z_{h}(0))^{2}A_h+z_{h}(1))^{2}q_0
-2z_{h}(0)z_{h}(1)B_h\right]}\nonumber\\
&\times\exp{\left[-\frac{1}{2C_\theta}(z_{\theta}(0))^{2}A_\theta
+z_{\theta}(1))^{2}q_0-2z_{\theta}(0)z_{\theta}(1)B_\theta\right]}
\\ 
\mbox{Prob}[\sigma(0),\sigma(1)|z_{h}(0),z_{\theta}(0)]&=
\mbox{Prob}(\sigma(0))\frac{\exp{\left(\beta\sigma(1)h(0)
+\beta\sigma^{2}(1)\theta(0)\right)}}{2\exp{\left(\beta\theta(0)\right)}\cosh(\beta
h(0))+1}\nonumber\\ 
&\times\frac{\exp{\left(\beta\sigma(2)h(1)
+\beta\sigma^{2}(2)\theta(1)\right)}}
{2\exp{\left(\beta\theta(1)\right)}\cosh(\beta
h(1))+1} 
\end{align} 
with the local fields at time $t=1$ given by 
\begin{eqnarray} 
h(1)&=&\frac{\xi}{a}m(1)+\frac{\alpha}{a^2}G_{h}(1,0)\sigma(0)
+\sqrt{\alpha}z_{h}(1)\\
\theta(1)&=&l(1)\eta+\frac{\alpha}{\tilde{a}^2}G_{h}(1,0)\sigma^{2}(0)
+\sqrt{\alpha}z_{\theta}(1)
\end{eqnarray} 
and the coeficients defined by 
\begin{eqnarray} 
A_{h}&\equiv&a^2q(1)+2aC_{h}(1,0)G_{h}(1,0)+q_{0}(G_{h}(1,0))^{2}\\ 
B_{h}&\equiv&a^2C_{h}(1,0)+aq_0G_h(1,0)\\ 
C_{h}&\equiv&q_0q(1)-C_{h}^2(1,0)\\ 
A_{\theta}&\equiv&\tilde{a}^2q(1)+2\tilde{a}C_{\theta}(1,0)
G_{\theta}(1,0)+q_{0}(G_{\theta}(1,0))^{2}\\
B_{h}&\equiv&\tilde{a}^2C_{\theta}(1,0)+\tilde{a}q_0G_\theta(1,0)\\ 
C_{h}&\equiv&q_0q(1)-C_{\theta}^2(1,0) 
\end{eqnarray} 

From these relations any quantity can be computed. We present the
final expressions for $m(2)$, $l(2)$ and $q(2)$. 
\begin{align} 
m(2)&=\frac{1}{a}\left\langle\!\left\langle\xi\sigma(2)
         \right\rangle\!\right\rangle\\
&=\frac{\tilde{a}}{2\pi\sqrt{A_{h}A_{\theta}}}\int
dz_{h}(1)dz_{\theta}(1)\exp{\left[-\frac{1}{2}
   \left(\frac{(z_{h}(1))^{2}}{A_{h}}+\frac{(z_{\theta}(1))^{2}}{A_{\theta}}
   \right)\right]}\nonumber\\
&\times\left\{AV_{\beta}(h^{++}(1),\theta^{++}(1))
   -CV_{\beta}(h^{-+}(1),\theta^{++}(1))
   +DV_{\beta}(h^{+0}(1),\theta^{+0}(1))\right.\nonumber\\
&\left.\quad-DV_{\beta}(h^{-0}(1),\theta^{+0}(1))
   +CV_{\beta}(h^{+-}(1),\theta^{++}(1))
   -AV_{\beta}(h^{--}(1),\theta^{++}(1))\right\}
\end{align} 
where the coeficients $A$, $C$ and $D$ coming from the distribution 
\begin{multline} 
\mbox{Prob}(\sigma(0),
\xi)=A\delta(\sigma(0)-1)\delta(\xi-1)
+B\delta(\sigma(0)-1)\delta(\xi) \\ 
+C\delta(\sigma(0)-1)\delta(\xi+1) 
+D\delta(\sigma(0))\delta(\xi-1) 
+E\delta(\sigma(0))\delta(\xi)\\
+D\delta(\sigma(0))\delta(\xi+1) 
+C\delta(\sigma(0)+1)\delta(\xi-1) \\
+B\delta(\sigma(0)+1)\delta(\xi) 
+A\delta(\sigma(0)+1)\delta(\xi+1)
\end{multline} 
defined by 
\begin{eqnarray} 
A&=&\frac{a}{4}(m_0+(1-a)l_0+q_0)\\ 
B&=&\frac{1-a}{2}(q_0-al_0)\\ 
C&=&\frac{a}{4}(-m_0+(1-a)l_0+q_0)\\ 
D&=&\frac{a}{2}(1-(1-a)l_0+q_0)\\ 
E&=&(1-a)(1+al_0-q_0) 
\end{eqnarray} 
The local field contributions appearing in these final equations have
the following form, due to the average over the initial conditions 
\begin{eqnarray} 
h^{uv}(1)&=&u\frac{m(1)}{a}+v\frac{\alpha}{a^2}
G_{h}(1,0)+\sqrt{\alpha}z_{h}(1)\\ 
\theta^{xy}(1)&=&\frac{x-a}{\tilde{a}}l(1)+y\frac{\alpha}{\tilde{a}^2}
G_{\theta}(1,0)+\sqrt{\alpha}z_{\theta}(1) 
\end{eqnarray} 
where $u,v\in\{-1,0,1\}$ and $x,y\in\{0,1\}$.

In the same way we find $q(2)$ and $l(2)$.

\end{document}